\documentclass[a4paper]{article}

\usepackage[utf8]{inputenc}
\usepackage[margin=1in]{geometry} 
\usepackage{cite}
\usepackage{amsmath,amssymb,amsfonts}
\usepackage{algorithm,algpseudocode}
\usepackage{graphicx}
\usepackage{textcomp}
\usepackage{url}
\usepackage{array}
\usepackage{xspace}
\usepackage{tablefootnote}
\usepackage{listings}
\usepackage{subfig}
\usepackage{color,soul}

\newcommand{\courier}{{\fontfamily{courier}\selectfont courier}\xspace}
\newcommand{\statistics}{{\fontfamily{courier}\selectfont statistics}\xspace}
\newcommand{\draw}{{\fontfamily{courier}\selectfont draw}\xspace}
\newcommand{\evoalglib}{{\fontfamily{courier}\selectfont evoalglib}\xspace}

\title{EISPY2D: An Open-Source Python Library for the Development and Comparison of Algorithms in Two-Dimensional Electromagnetic Inverse Scattering Problems\footnote{This work has been submitted to the IEEE for possible publication. Copyright may be transferred without notice, after which this version may no longer be accessible.}}
\author{Andr\'e Costa Batista\textsuperscript{1}, Ricardo Adriano\textsuperscript{2}, and Lucas S. Batista\textsuperscript{2}}
\date{\small \textsuperscript{1}Graduate Program in Electrical Engineering\\Universidade Federal de Minas Gerais, Belo Horizonte 31270-901, Brazil\\E-mail: andre-costa@ufmg.br\\\textsuperscript{2}Department of Electrical Engineering\\Universidade Federal de Minas Gerais, Belo Horizonte 31270-901, Brazil\\E-mail: lusoba@ufmg.br; rluiz@ufmg.br}

\begin{document}

	\maketitle

	\begin{abstract}
		Microwave Imaging is an essential technique for reconstructing the electrical properties of an inaccessible medium. Many approaches have been proposed employing algorithms to solve the Electromagnetic Inverse Scattering Problem associated with this technique. In addition to the algorithm, one needs to implement adequate structures to represent the problem domain, the input data, the results of the adopted metrics, and experimentation routines. We introduce an open-source Python library that offers a modular and standardized framework for implementing and evaluating the performance of algorithms for the problem. Based on the implementation of fundamental components for the execution of algorithms, this library aims to facilitate the development and discussion of new methods. Through a modular structure organized into classes, researchers can design their case studies and benchmarking experiments relying on features such as test randomization, specific metrics, and statistical comparison. To the best of the authors' knowledge, it is the first time that such tools for benchmarking and comparison are introduced for microwave imaging algorithms. In addition, two new metrics for location and shape recovery are presented. In this work, we introduce the principles for the design of the problem components and provide studies to exemplify the main aspects of this library. It is freely distributed through a Github repository that can be accessed from \url{https://andre-batista.github.io/eispy2d/}.
	\end{abstract}

	\hfill\break
	\textbf{Keywords}: Microwave Imaging, Electromagnetic Inverse Scattering Problem, Open-Source Library, Optimization, Comparison of Algorithms.

	\section{Introduction}\label{sec:introduction}
	
	Microwave Imaging (MWI) is a technique whose potential began to be investigated in the 1970s \cite{shan1979microwave} and has attracted much research since then \cite{nikolova2017introduction,pastorino2010ch11}. Its main objective is the image reconstruction of the interior of a region of interest through the investigation of its electrical properties. These properties are investigated through the propagation of electromagnetic waves in the spectrum between 300 kHz and 300 GHz. Because of their advantages, such as the relatively low cost of the necessary equipment and the use of non-ionizing radiation \cite{pastorino2010ch10}, many researchers have developed techniques for different types of applications, including breast  \cite{conceicao2016introduction,aldhaeebi2020review} and brain stroke cancer detection \cite{ireland2011microwave}, non-destructive testing and evaluation \cite{kharkovsky2007microwave}, through-wall imaging \cite{baranoski2008through}, among others. The main challenge is the solution of the Electromagnetic Inverse Scattering Problem (EISP) \cite{chen2018computational} which is ill-posed, non-linear, and multi-modal, i.e., it can have multiple local minima when formulated as an optimization problem.
	
	Many methodologies have already been proposed in the literature to solve EISPs, mainly in their two-dimensional formulation. Based on common characteristics, these methodologies can be classified in qualitative and quantitative or deterministic and stochastic \cite{pastorino2010ch5,pastorino2010ch6,pastorino2010ch7}. Qualitative methods are those whose interest is to determine the position and shape of the scatterers (e.g., Linear Sampling \cite{colton2003linear} and Level-Set \cite{shah2018fast}), while quantitative methods further determine the value of the electrical properties of the medium (e.g., Born Iterative Method \cite{wang1989iterative,batista2021quadratic} and Contrast Source Inversion \cite{berg1997contrast}). Deterministic methods are those that determine the solution based on a set of fixed steps and, therefore, always obtain the same solution for the same input (e.g., Distorted Born Iterative Method \cite{chew1990reconstruction}). On the other hand, stochastic methods define theirs steps based on pseudo-random operations to determine the solution and, therefore, they can obtain different solutions for the same input (e.g., Evolutionary Algorithms \cite{rocca2009evolutionary}). Other recent methods that have gained attention in the literature are the Subspace Optimization Method (SOM) \cite{chen2010subspace} and the regularization in Lebesgue spaces with variable exponents \cite{estatico2012novel}. Furthermore, the application of Deep Learning methods has attracted a lot of community consideration as these techniques can speed up image recovery and facilitate real-time imaging \cite{chen2020review}. 
	
	The evaluation of the proposed methodologies is not an easy task. There are many parameters in the problem configuration that can influence the performance of the algorithm, such as the contrast level of the scatterers, their amount, their distribution in space, their geometry, the noise level present in the data, the number of measurements and sources, among others. When the algorithms are tested with known solutions, the quality of the reconstructions is usually quantified by the relative-mean-square-error of the electric property estimation in each image pixel or in the scatterer regions (see chapter 6 of \cite{nikolova2017introduction}), which does not necessarily reflect the quality of the scatterer geometry reconstruction and the accuracy of its position. However, in the context of image reconstructions of the interior of the breast, there is a package with adequate tools to assess the quality of detection and reconstruction of tumors through a set of indicators \cite{kurrant2021evaluating,kurrant2021mwsegeval}. 
	
	Neither is the comparison among methods an easy task as well. In addition to the algorithms being developed on custom platforms, the codes are generally not available. Because of this, there are few works whose objective was to compare state-of-the-art algorithms (e.g., \cite{mohhaddam1991comparison} and \cite{gilmore2009comparison}), and new methodologies are usually tested against small variations in their choices or traditional methods (e.g. \cite{kuiwen2018hybrid,estatico2018quantitative,salucci2017multifrequency}). Although the use of real-world data is a relevant feature when experimenting with methodologies, the low number of test problems can prevent an evaluation of the average performance of the methods. This average measure can be important information both for applications that require a specific quality level in the reconstruction and for statistically inferring the superiority of a method, especially when it depends on stochastic processes \cite{beiranvand2017best}. 
	
	Given this gap in the literature, we propose an open-source library that implements a platform whose objective is to facilitate the development and comparison of algorithms for EISP, considering the two-dimensional case. This library was implemented in Python and its project repository is available at \url{https://andre-batista.github.io/eispy2d/}. This platform, called \textbf{eispy2d}, is based on a standard definition of the problem geometry and allows the exploration of different configurations. In addition to the possibility of synthesizing tests for case studies or benchmarks based on the definition of some control parameters, the platform has a set of quality indicators gathered from the literature and two new specific ones to quantify the quality of the location and shape recovery of the scatterers. Furthermore, adequate statistical tools were implemented for comparing the results either between stochastic methods in case studies or benchmarks considering deterministic and stochastic algorithms. In this sense, the originality of this work consists both in the elaboration of a common platform for EISPs and the application of more robust techniques to compare these methods. Moreover, the two new metrics for shape and position retrievement are also novelties. This work contributes to the state-of-the-art literature as it introduces more solid practices for evaluating the performance of the algorithms proposed for the problem and a simpler and general approach to measure location and shape error.
	
	This paper describes the library, its structure, functionality and introduces what is necessary for a user to develop their method in \textbf{eispy2d} as well as test and evaluate it. First, the two-dimensional formulation of the EISP, that will be adopted in the library, is presented and the definition of spatial geometry, constants, known and unknown variables, equations, and the types of allowed materials are also highlighted (Section \ref{sec:eisp}). Then, in Section \ref{sec:library}, the library is detailed, presenting its main modules and classes. This description serves as a reference for the main features of the platform, as well as a starting point for future contributions. In Section \ref{sec:example}, three examples that illustrate the use of the library are provided: (i) the manual definition of a single problem and its solution; (ii) the definition of a case study and the analysis of its results; and (iii) the definition of a benchmark as well as the analysis of the results of the involved algorithms. Finally, in Section \ref{sec:conclusion}, we conclude the article by pointing out features that can be implemented in the future.
	
	\section{The Electromagnetic Inverse Scattering Problem}\label{sec:eisp}
	
	When an electromagnetic wave, generated by a known impressed source, propagates through free space and impinges on a medium with different electromagnetic properties, part of the field is transmitted to the interior of the medium and another part is reflected. Thus, the total field observed at a point in free space can be interpreted as the sum of the field radiated by the impressed source (incident field) and the perturbation resulting from the presence of an object (scattered field). To describe the basic configuration of an EISP, we need to start by defining the region where the scatterers are located and the one where the total field is observed. Assuming a two-dimensional Cartesian space where the unit of spatial coordinates is expressed in meters, the region where the scatterers are located will be called $D$\footnote{This region is also commonly referred to in the literature as Region of Interest (ROI) or Domain of Interest (DOI).} and will be defined as a rectangle of $L_x$ and $L_y$ dimensions centered on the origin of the coordinate axis, i.e.,
	\begin{equation}
		D = \left\{(x, y)~|~x\in[-L_x/2, L_x/2]\mathrm{~and~}y\in[-L_y/2,L_y/2]\right\}.  \label{eq:D}
	\end{equation}
	
	The region where the total field can be observed will be defined as a circle centered at the origin of the plane whose radius will be called $R_O$. Therefore, this region, called $S$, can be defined as:
	\begin{equation}
		S = \left\{(x, y) ~|~ x = R_O\cos\theta \mathrm{~and~} y = R_O\sin\theta, ~\forall~ \theta\in[0, 2\pi] \right\}. \label{eq:S}
	\end{equation}
	
	\begin{figure}
		\centering
		\includegraphics[width=3.5in]{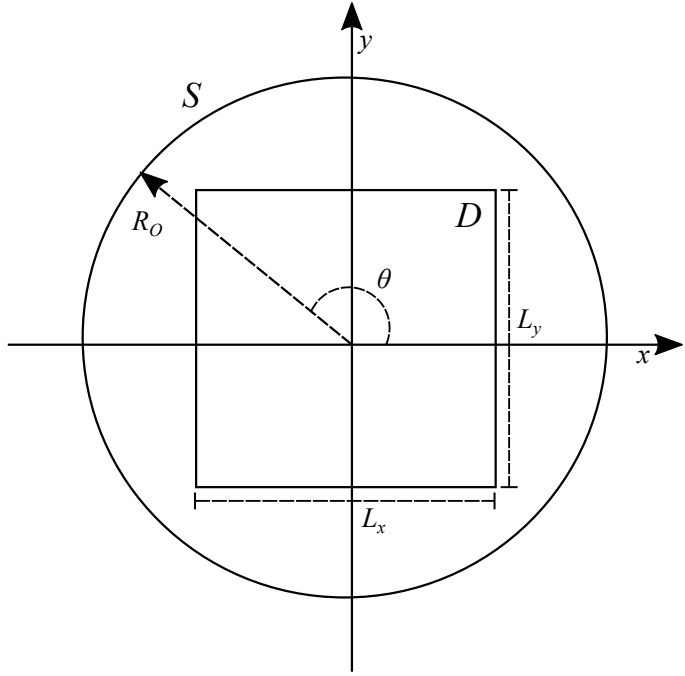}
		\caption{Geometry definition.}
		\label{fig:geometry}
	\end{figure}
	
	Regions $D$ and $S$ are illustrated in Fig. \ref{fig:geometry}. A point in spaces $D$ and $S$ will be represented by $\boldsymbol{\rho}=\langle x, y \rangle$. The fields will be represented in the frequency domain assuming the time-factor $e^{j\omega t}$ (where $j$ is the imaginary unit, $\omega$ is the angular frequency expressed in radians per seconds, and $t$ is the time in seconds). If the Transverse Magnetic mode in $z$ (TMz) for wave polarization and the Silver-M\"uller radiation boundary condition \cite{pastorino2010ch2} are assumed, then the equation that relates the total electric field $E_z$ (V/m), the incident electric field $E^i_z$ (V/m), and the contrast function $\chi$ that maps the electrical properties of the medium, is expressed by \cite{chew1995waves}:
	\begin{align}
		E_z(\boldsymbol{\rho}) &= E^i_z(\boldsymbol{\rho})  \nonumber\\ &- \frac{jk_b^2}{4} \int_D dS^\prime H^{(2)}_0(k_b|\boldsymbol{\rho}-\boldsymbol{\rho^\prime}|)\chi(\boldsymbol{\rho^\prime})E_z(\boldsymbol{\rho^\prime}),~ \boldsymbol{\rho}\in S, \label{eq:integralequation}
	\end{align}
	
	\noindent where $k_b=\sqrt{\mu_0\epsilon_b}$ is the wavenumber (m) referring to the homogeneous background medium in which the incident wave propagates, $\epsilon_b$ is the permittivity of the background medium (F/m), $dS^\prime$ is an infinitesimal element of area in $D$, $H^{(2)}_0$ is the Hankel function of second kind  and zero order and the contrast function is expressed by:
	\begin{equation}
		\chi(\boldsymbol{\rho}) = \frac{\epsilon_r(\boldsymbol{\rho})}{\epsilon_{rb}} - 1 - j\frac{\sigma(\boldsymbol{\rho})-\sigma_b}{\omega\epsilon_b} ,\label{eq:contrastfunction}
	\end{equation}
	
	\noindent where $\epsilon_r$ is the relative permittivity, $\epsilon_{rb}$ is the relative permittivity of the background medium, $\sigma$ is the conductivity (S/m), and $\sigma_b$ is the conductivity of the background medium (S/m). Therefore, the model assumes linear, isotropic, non-dispersive, non-magnetic materials and homogeneous background medium. 
	
	Since the scattered field, $E^s_z$ (V/m), can be written as the difference between the total and the incident fields, i.e., $E^s_z=E_z-E^i_z$, then \eqref{eq:integralequation} can be rewritten as:
	\begin{equation}
		E^s_z(\boldsymbol{\rho}) = -\frac{jk_b^2}{4} \int_D dS^\prime H^{(2)}_0(k_b|\boldsymbol{\rho}-\boldsymbol{\rho^\prime}|)\chi(\boldsymbol{\rho^\prime})E_z(\boldsymbol{\rho^\prime}),~ \boldsymbol{\rho}\in S, \label{eq:dataequation}
	\end{equation}
	
	\noindent which is known as the \textit{data equation}. Another very common one is the \textit{state equation} expressed by:
	\begin{align}
		E_z(\boldsymbol{\rho}) &= E^i_z(\boldsymbol{\rho}) \nonumber\\ & -\frac{jk_b^2}{4} \int_D dS^\prime H^{(2)}_0(k_b|\boldsymbol{\rho}-\boldsymbol{\rho^\prime}|)\chi(\boldsymbol{\rho^\prime})E_z(\boldsymbol{\rho^\prime}),~ \boldsymbol{\rho}\in D. \label{eq:stateequation}
	\end{align}
	
	In an EISP, the scattered field is known in a finite set of $N_M$ equidistant points in $S$, sampled for each illumination of a finite set of $N_S$ incident sources. Therefore, the input data are the $N_MN_S$ values of $E^s_z$ in $S$ and the incident field in $D$ for each of the $N_S$ impressed sources on the known homogeneous background medium. Unknown variables that are determined by the methods are $\chi$ and $E_z$ in $D$.
	
	It is important to highlight that other formulations besides  \eqref{eq:dataequation} and \eqref{eq:stateequation} are possible. A very traditional one is the contrast-source formulation in which \eqref{eq:dataequation} and \eqref{eq:stateequation} are rewritten to determine $\chi$ and the equivalent current $J^{eq}_z=\chi E_z$ induced in the scatterer region. There are also other modifications recently proposed in the literature that reduce the Degree of Non-Linearity (DNL) of the problem \cite{bevacqua2021quantitative}.Alternatively, the linear combinations of the equations, called Virtual Experiments, have demonstrated the potential to make the inversion task simpler \cite{bevacqua2021simple}.
	
	Finally, the justifications for the choices in modeling the problem are listed below:
	\begin{itemize}
		\item We decided to approach the two-dimensional case since the problem is less complex and, therefore, new ideas are usually tested first in this case.
		\item We decided to address only linear, isotropic, non-dispersive, and non-magnetic materials because most methodologies are implemented considering this configuration. Dispersive materials are generally treated in specific applications.
		\item We chose to represent EISPs with only one operating frequency to make it easier to describe the problem in terms of wavelength and because any method that uses more frequencies is expected to perform better as the problem becomes less ill-posed.
	\end{itemize}
	
	\section{The eispy2d Library}\label{sec:library}
	
	The \textbf{eispy2d} library is a Python implementation of the necessary framework to represent EISPs following the formulation described in the previous section, to develop specialized algorithms for the problem, and to analyze and compare results. Our goals with this library are:
	\begin{itemize}
		\item To provide a structure capable of representing instances of two-dimensional EISPs with the necessary information, thus saving researchers' time concerning the implementation of the necessary resources to test their new ideas.
		\item To develop a common platform for method development equipped with frequently employed tools.
		\item To implement an adequate structure for generating, analyzing, and comparing results, equipped with tools for test randomization, quality measurement, and statistical inference.
	\end{itemize}
	
	To achieve these goals, the implementation of this platform was guided by the following design decisions:
	\begin{itemize}
		\item We chose the Python language because it is open-source and represents a very popular language with many free packages available.
		\item We chose to follow the Object-Oriented Programming paradigm. Therefore, the information will be organized into classes with attributes and methods.
		\item To facilitate the execution of the algorithms, we use the concept of inheritance to create the outline of the methods (forward, inverse, deterministic, and stochastic) using abstract classes. Thus, new methodologies will have to be implemented as derived classes keeping the input and output pattern in some functions.
	\end{itemize}
	
	Regarding the implementation, class names are written in italics throughout the text. Class attribute and function names are written in quotation marks. The modules, i.e., files that contain a set of tools for some purpose, are written in the \courier font.
	
\begin{figure*}
		\centering
		\includegraphics[width=\textwidth]{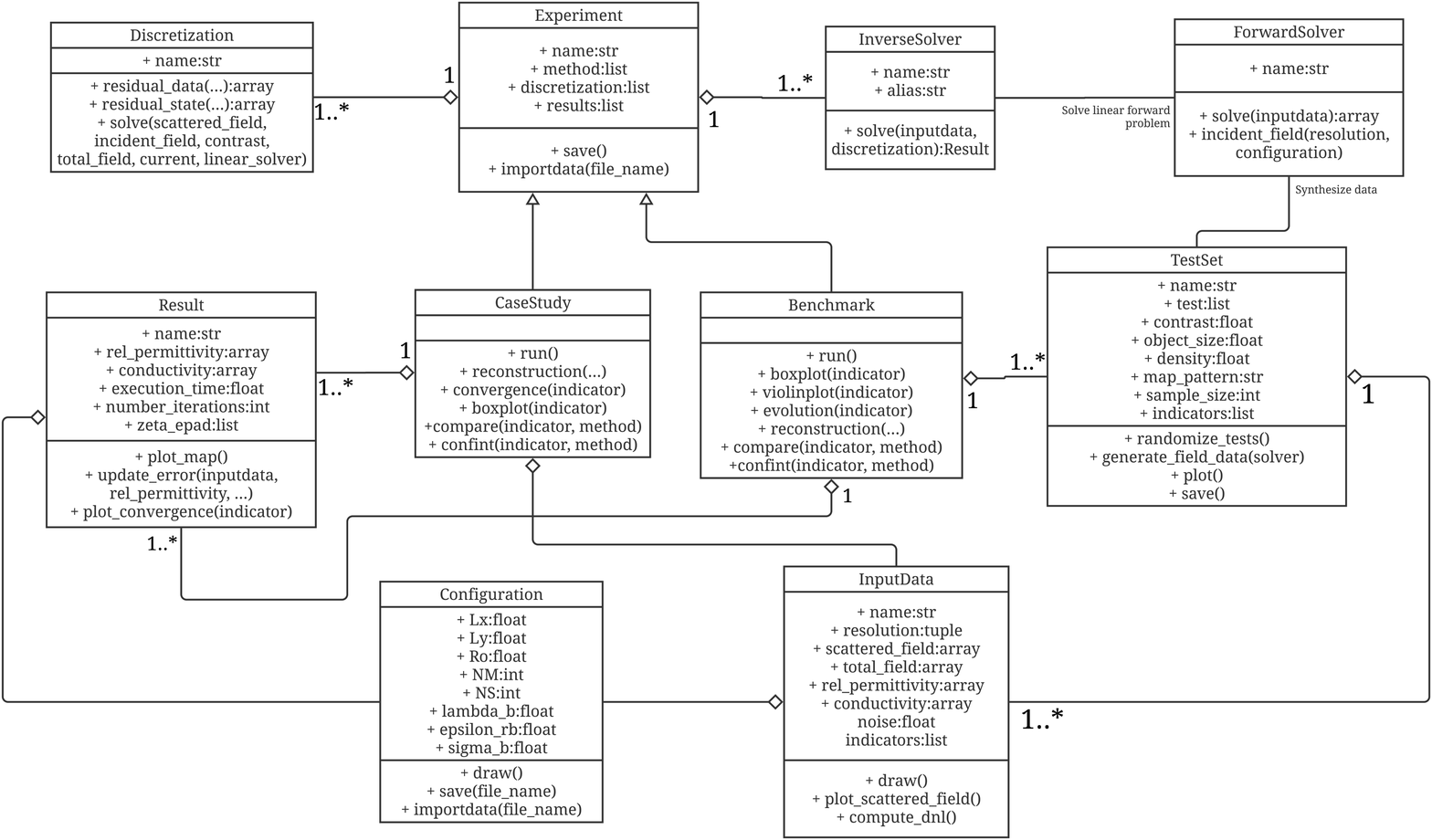}
		\caption{\textbf{eispy2d} main structure  UML Class Diagram.}
		\label{fig:uml:general}
	\end{figure*}

	The general structure of \textbf{eispy2d} is represented through the UML class diagram in Fig.\ref{fig:uml:general}. An EISP test problem is represented by the \textit{Configuration} and \textit{InputData} classes which contain the domain definition and input parameters of the problem, respectively. The abstract \textit{InverseSolver} class represents any method that solves the non-linear inverse problem, and it generates an object of the \textit{Result} class containing the results of the execution. The abstract class \textit{Experiment} contains the general attributes of any form of experiment (case study or benchmarking). The \textit{TestSet} class implements the test randomization process based on the definition of some control parameters. The abstract \textit{Discretization} class represents a general scheme for the discretization of equations. Finally, the \textit{ForwardSolver} class implements a general structure of algorithms that determine the electric field based on some medium configuration and is employed either to synthesize the input data of experiments or inside inversion methods that depend on the solution of the direct problem.
	
	In the next subsections, each of these classes are detailed, their roles are explained, and derived classes are introduced as well as other modules that complement the library. A summary is presented in Table \ref{tab:summary}.
	
	\begin{table*}
		\centering
		\caption{List of classes and modules in the library. Classes are written in \textit{italics} while modules are written in \courier font.}
		\begin{tabular}{m{0.25\textwidth}m{0.25\textwidth}m{0.25\textwidth}m{0.1\textwidth}}
			\hline
			Classes and modules & Role & Derivations, contends, user parameters & Section \\\hline
			\textit{Configuration} & Represent the problem domain & $L_x$, $L_y$, $N_M$, $N_S$, $\lambda_b$, $\epsilon_{rb}$, $\sigma_b$ & \ref{subsub:configuration} \\
			\textit{Discretization} & Methods for Discretization of Equations& \textit{Collocation}, \textit{Richmond} &  \ref{subsub:discretization} \\
			\draw & Routines for drawing geometric figures & triangle, circle, etc & \ref{subsub:draw} \\
			\textit{Experiment} & Represent experiments & \textit{CaseStudy}, \textit{Benchmark} &  \ref{sub:experiment} \\
			\evoalglib & Framework for exploring evolutionary algorithms & \textit{DE}, \textit{PSO}, \textit{GA}, \textit{Representation}, etc & \ref{subsub:evoalglib} \\
			\textit{ForwardSolver} & Forward problem solvers & \textit{MoM\_CG\_FFT} & \ref{subsub:forward} \\
			\textit{InputData} & Store input data for forward and inverse solvers & Scattered field, relative permittivity map, etc & \ref{subsub:inputdata} \\
			\textit{InverseSolver} & Inverse problem solvers & \textit{Deterministic}, \textit{Stochastic}, \textit{BIM}, \textit{EvolutionaryAlgorithm} & \ref{subsub:inverse}  \\
			\textit{Regularization} & Regularization methods for linear inverse problems & \textit{Tikhonov}, \textit{Landweber}, \textit{ConjugatedGradient}, \textit{SpectralCutOff} & \ref{subsub:regularization} \\
			\textit{Result} & Store results of a single execution & Recovered images, number of iterations, indicators, etc & \ref{subsub:result} \\
			\statistics & Routines for statistical tests & Normality plots, Multiple samples test, etc & \ref{subsub:statistics} \\
			\textit{StopCriteria} & Check stop criteria & Number of iterations, maximum iterations without improvement, etc & \ref{subsub:stopcriteria} \\
			\textit{TestSet} & Represent a test set for benchmark purposes & $\chi_{max}$, $r_{max}$, sample size, etc & \ref{subsub:testset} \\
			\hline
		\end{tabular}
		\label{tab:summary}
	\end{table*}
	
	\subsection{Basic Classes}\label{sub:basic}
	
	Three basic classes have been defined to gather input and output information from the algorithms. They are responsible for storing the data required to describe the problem and the performance of the methods. In addition to archiving the information, routines for viewing the collected data are also offered.
	
	\subsubsection{\textit{Configuration}}\label{subsub:configuration}
	
	The \textit{Configuration} class is responsible for storing the value of the parameters that describe the problem domain and the background medium in which the incident wave propagates. Therefore, all constants introduced in Section \ref{sec:eisp} are defined in this class, i.e., $L_x$, $L_y$, $N_M$, $N_S$, $R_O$, $\epsilon_{rb}$, and $\sigma_b$. Basic information about the incident waveform configuration is also stored in this class, i.e., wave magnitude $E_0$ (V/m), linear frequency $f$ (Hz), wavelength $\lambda_b$ (m), and wavenumber $k_b$ (1/m). The user will have to specify, in addition to $E_0$, either $f$ or $\lambda_b$ when constructing an object. From one of these constants, the other is automatically calculated just like $k_b$. It is important to note that $L_x$, $L_y$, and $R_O$ can be specified both in meters and as a function of $\lambda_b$. Also, if $N_M$ or $N_S$ are not specified, they are automatically set to triple the number of Degrees of Freedom (DOF) of the problem \cite{bucci1997electromagnetic}. In these cases, the radius of the scatterer, necessary information for calculating DOF, is defined as $\min(L_x/2, L_y/2)$. That is, the DOF is calculated to cover the worst case, which is a scatterer whose diameter is the size of $D$. The choice for triple the value was made from preliminary studies of observation of the scattered field and the reconstruction quality of the Born Iterative Method (BIM) with different proportions of DOF regarding scenarios of strong and weak scatterers. Finally, it is highlighted that this class also contains two Boolean variables that can be used to assume a scenario where only perfect dielectric scatterers ($\sigma=0$) or good conductors ($\epsilon_r=1$) are considered. When any of these variables is true, the methods can eliminate, respectively, the imaginary or real part of \eqref{eq:contrastfunction}.
	
	\subsubsection{\textit{InputData}}\label{subsub:inputdata}
	
	The \textit{InputData} class is used as one of the input arguments of the forward or inverse solvers. Therefore, it can store information such as the scattered field data, the total field data, the relative permittivity and conductivity images, the noise level of the field data, and the set of performance indicators that must be considered. An object of the \textit{Configuration} class is also stored as an attribute of that class. Therefore, if an object of this class is used as input to a forward resolver, it is expected to contain, in addition to the domain configuration data, at least the image of the permittivity or conductivity. In the case of use as an input to an inverse solver, it is necessary that, in addition to the domain configuration data, at least the $E^s_z$ data are present. If the list of indicators contains one that measures the error concerning the original images of $\chi$ or $E_z$, then the relative permittivity and conductivity and total field image data must be present. Another attribute of the class is the resolution of the images involved in the problem. This attribute is specified as a tuple ($N_y$, $N_x$) where $N_x$ and $N_y$ denote the number of pixels on the image's $x$ and $y$ axes, respectively. The standardization of the format of the matrices that contain the field information and the electrical properties is described in Table \ref{tab:inputdata}. Finally, it is worth mentioning that the DNL associated with the inverse problem can be calculated and stored in the object. The implementation of the DNL calculation follows the traditional model ($H_0$ \cite{bevacqua2021quantitative}) based on \eqref{eq:dataequation} and calculated from Richmond discretization \cite{richmond1965scattering}.
	
	\begin{table}
		\centering
		\caption{Dimensions of the arrays that store the data present in an object of the \textit{InputData} class.}
		\begin{tabular}{cccc}
			\hline
			Data & Unit & Variable type & Array format \\\hline
			Scattered Field & (V/m) & Complex & $N_M \times N_S$ \\
			Total Field\tablefootnote{The image of the total field associated with each of the N sources is present in each column of the matrix and the pixels are ordered according to the following pattern: the last axis index changing fastest, back to the first axis index changing slowest (C-like).} & (V/m) & Complex & $N_yN_x \times N_S$ \\
			Relative Permittivity & & Float & $N_y \times N_x$ \\
			Conductivity & (S/m) & Float & $N_y \times N_x$ \\\hline
		\end{tabular}
		\label{tab:inputdata}
	\end{table}
	
	\subsubsection{\textit{Result}}\label{subsub:result}
	
	The \textit{Result} class is responsible for storing the output data of inverse methods. So it has attributes to store the total field and electrical properties images. In addition, it stores each performance indicator as a list where each element references an iteration of the inverse solver. It must also contain an object of the \textit{Configuration} class to support the calculation. In addition to the indicators, the number of iterations, the execution time (in seconds), the number of evaluations, and the objective function value can also be stored (the last two are more common in stochastic algorithms).
	
	The set of implemented indicators aims to evaluate, in different ways, the error in estimating the total field, the electrical properties, and the corresponding scattered field. These indicators are equal or equivalent to those found in the literature \cite{wang1989iterative,salucci2017multifrequency,berg1997contrast}. However, two new indicators were also proposed: one for the error in the location of the scatterers and the other for the error in the reconstruction of the scatterer shapes. Below, we present a list of each of the 13 indicators\footnote{Each variable with the superscript ``$^g$'' means the ground-truth value. On the other hand, the ones with the superscript ``$^r$'' mean the recovered values or computed through the recovered values.}. These are also summarized in Table \ref{tab:result}.
	
	\begin{itemize}
		\item Residual Norm ($\zeta_{rn}$): error norm between the scattered field input data and the one corresponding to the total field and contrast information retrieved by the algorithm\footnote{Regarding the Residual Norm, the operator ``$\mathrm{inner}$'' is defined as the product of the argument with its complex conjugate value. Furthermore, it is assumed that the scattered field, in addition to varying with the measurement angle $\theta$, varies according to either the angle of inclination of the incident wave (in the case of infinite plane waves) or the position angle of the source (in the case of current sources printed), both cases represented by the variable $\phi$. Finally, the Trapezoid Rule is applied to estimate the integral since we have only sampled values of the field.}.
		\begin{align}
			\zeta_{RN} = \left(\int_0^{2\pi} \right. & \left. \int_0^{2\pi} \mathrm{inner}(E^{s,g}_z(\theta,\phi) \right. \nonumber \\
			&  -E^{s,r}_z(\theta,\phi))d\theta d\phi \bigg)^{1/2} \mathrm{~~(V/m)} \label{eq:zeta:rn}
		\end{align}
		\item Percentage Average Deviation of Residues ($\zeta_{rpad}$): percentage average error between the scattered field given as input and the one corresponding to the solution estimated by the algorithm.
		\begin{align}
			\zeta_{RPAD} &= 100\times\frac{1}{2N_MN_S} \nonumber \\
			& \times \sum\limits_{m=1}^{N_M}\sum\limits_{s=1}^{N_S} \left( \left| \frac{\Re\{E^{s,g}_{z,ms}-E^{s,r}_{z,ms}\}}{\Re\{E^{s,g}_{z,ms}\}} \right|  \right. \nonumber \\
			& \left. +  \left| \frac{\Im\{E^{s,g}_{z,ms}-E^{s,r}_{z,ms}\}}{\Im\{E^{s,g}_{z,ms}\}} \right| \right) \mathrm{~~(\%/sample)} \label{eq:zeta:rpad}
		\end{align}
		\item Percentage Average Deviation of Relative Permittivity Estimation ($\zeta_{\epsilon PAD}$): Average percent error per pixel in estimating relative permittivity over $D$.
		\begin{equation}
			\zeta_{\epsilon PAD} = 100 \times \frac{1}{N_xN_y} \sum\limits_{i=1}^{N_x}\sum\limits_{j=1}^{N_y} \left| \frac{\epsilon^{g}_{r,ij}-\epsilon^{r}_{r,ij}}{\epsilon^{g}_{r,ij}} \right| \mathrm{~~(\%/pixel)} \label{eq:zeta:epad}
		\end{equation}
		\item Percentage Average Deviation of Relative Permittivity Estimation in Scatterer Regions ($\zeta_{\epsilon OE}$): average percent error per pixel in estimating relative permittivity only in regions whose contrast is different from zero in the ground-truth image\footnote{The set of pixels, which are within the scatterer region in the original image, is represented by the $\mathbf{N^{\chi\neq0}}$ symbol.}.
		\begin{equation}
			\zeta_{\epsilon OE} = 100 \times \frac{1}{|\mathbf{N^{\chi\neq0}}|} \sum\limits_{n\in\mathbf{N^{\chi\neq0}}} \left| \frac{\epsilon^{g}_{r,n}-\epsilon^{r}_{r,n}}{\epsilon^{g}_{r,n}} \right| \mathrm{~~(\%/pixel)} \label{eq:zeta:eoe}
		\end{equation}
		\item Percentage Average Deviation of Relative Permittivity Estimation in Background Regions ($\zeta_{\epsilon BE}$): average percent error per pixel in estimating relative permittivity only in regions whose contrast is equal to zero in the ground-truth image\footnote{The set of pixels, which are within the background region in the original image, is represented by the $\mathbf{N^{\chi=0}}$ symbol.}.
		\begin{equation}
			\zeta_{\epsilon BE} = 100 \times \frac{1}{|\mathbf{N^{\chi=0}}|} \sum\limits_{n\in\mathbf{N^{\chi=0}}} \left| \frac{\epsilon^{g}_{r,n}-\epsilon^{r}_{r,n}}{\epsilon^{g}_{r,n}} \right| \mathrm{~~(\%/pixel)} \label{eq:zeta:ebe}
		\end{equation}
		\item Average Conductivity Deviation ($\zeta_{\sigma AD}$): average percent error per pixel in the conductivity estimate over $D$.
		\begin{equation}
			\zeta_{\sigma AD} = \frac{1}{N_xN_y} \sum\limits_{i=1}^{N_x}\sum\limits_{j=1}^{N_y} \left| \sigma^{g}_{ij}-\sigma^{r}_{ij} \right| \mathrm{~~(S/pixel)} \label{eq:zeta:sad}
		\end{equation}
		\item Average Conductivity Deviation in Scatterer Regions ($\zeta_{\sigma OE}$): average percent error per pixel in the conductivity estimate only in regions whose contrast is different from zero in the ground-truth image.
		\begin{equation}
			\zeta_{\sigma OE} = \frac{1}{|\mathbf{N^{\chi\neq0}}|} \sum\limits_{n\in\mathbf{N^{\chi\neq0}}} \left| \sigma^{g}_{n}-\sigma^{r}_{n} \right| \mathrm{~~(S/pixel)} \label{eq:zeta:soe}
		\end{equation}
		\item Average Conductivity Deviation in Background Regions ($\zeta_{\sigma BE}$): average percent error per pixel in the conductivity estimate only in regions whose contrast is equal to zero in the ground-truth image.
		\begin{equation}
			\zeta_{\sigma BE} = \frac{1}{|\mathbf{N^{\chi=0}}|} \sum\limits_{n\in\mathbf{N^{\chi=0}}} \left| \sigma^{g}_{n}-\sigma^{r}_{n} \right| \mathrm{~~(S/pixel)} \label{eq:zeta:sbe}
		\end{equation}
		\item Average Percentage Deviation in Total Field Magnitude Estimation ($\zeta_{TFMPAD}$): average percent error per pixel in estimating the magnitude of the total field in $D$.
		{\footnotesize
			\begin{align}
				\zeta_{TFMPAD} = & 100 \times \frac{1}{N_xN_yN_S} \nonumber \\ & \times \sum\limits_{i=1}^{N_x}\sum\limits_{j=1}^{N_y}\sum\limits_{s=1}^{N_S} \left| \frac{|E^{g}_{z,ijs}|-|E^{r}_{z,ijs}|}{|E^{g}_{z,ijs}|} \right| \mathrm{~~(\%/pixel)} \label{eq:zeta:tfmpad}
		\end{align}}
		\item Average Deviation in Total Field Phase Estimation ($\zeta_{TFPAD}$): average percent error per pixel in estimating the total field phase in $D$.
		{\footnotesize
			\begin{align}
				\zeta_{TFPAD} = & \frac{1}{N_xN_yN_S} \nonumber \\ & \times \sum\limits_{i=1}^{N_x}\sum\limits_{j=1}^{N_y}\sum\limits_{s=1}^{N_S} \left| \angle E^{g}_{z,ijs}-\angle E^{r}_{z,ijs} \right| \mathrm{~~(rad/pixel)} \label{eq:zeta:tfpad}
		\end{align}}
		\item Total variation: total variation of the contrast image.
		\begin{equation}
			\zeta_{TV} = \int_D \frac{|\nabla\chi|^2}{|\nabla\chi|^2+1} dS \label{eq:zeta:tv}
		\end{equation}
		\item Position Error ($\zeta_P$): error in detecting the position of the scatterers. This indicator is based on the radius between the scatterer distribution center in the ground-truth image and the recovered one. Since the coordinates of the centers are calculated in terms of the fraction of the domain length, the indicator represents the percentage displacement of the scatterers in the reconstructed image to the ground-truth one. It is worth noting that this indicator can be applied in cases with one or multiple scatterers. The methodology to calculate this indicator is described in Alg.\ref{alg:zeta:p}.
		\item Shape Error ($\zeta_{S}$): error in the shape reconstruction of the scatterers. This indicator is defined in terms of the ratio between the area of false-negative and false-positive regions in the recovered image and the area of the scatterers in the ground-truth one. Therefore, when this ratio is multiplied by 100, it measures the percentage difference between the reconstructed and the original image concerning the latter. The justification for this definition is that this indicator tends to zero when the shapes of the scatterers are the same in both images, and it can be applied in scenarios with multiple scatters. To isolate the effect of the scatterer placement error on the retrieved image, the same scatterer distribution center calculation procedure in Alg.\ref{alg:zeta:p} is applied to center the images and allow a fairer comparison of the scatterer shape\footnote{The description of the procedure for calculating $\zeta_{S}$ and $\zeta_{S}$ is available at \url{https://github.com/andre-batista/eispy2d/blob/main/demo/position_and_shape_metrics.ipynb}}.
	\end{itemize}
	
	\begin{table*}
		\centering
		\caption{Summary of performance indicators implemented in the \textit{Result} class.}
		\begin{tabular}{m{.25\textwidth}m{.1\textwidth}m{.1\textwidth}m{.08\textwidth}c}
			\hline
			Name & Symbol & Attribute & Unit & Formula \\\hline
			Residual Norm & $\zeta_{RN}$ & ``zeta\_rn'' & (V/m) & \eqref{eq:zeta:rn} \\
			Percentage Average Deviation of Residual Norm & $\zeta_{RPAD}$ & ``zeta\_rpad'' & (\%/sample) &  \eqref{eq:zeta:rpad} \\
			Percentage Average Deviation of Relative Permittivity & $\zeta_{\epsilon PAD}$ & ``zeta\_epad'' & (\%/pixel) & \eqref{eq:zeta:epad} \\
			Percentage Average Deviation of Relative Permittivity at scatterer areas & $\zeta_{\epsilon OE}$ & ``zeta\_eoe'' & (\%/pixel) & \eqref{eq:zeta:eoe} \\
			Percentage Average Deviation of Relative Permittivity at background areas & $\zeta_{\epsilon BE}$ & ``zeta\_ebe'' & (\%/pixel) & \eqref{eq:zeta:ebe} \\
			Average Deviation of Conductivity & $\zeta_{\sigma AD}$ & ``zeta\_sad'' & (S/pixel) & \eqref{eq:zeta:sad} \\
			Average Deviation of Conductivity at scatterer areas & $\zeta_{\sigma OE}$ & ``zeta\_soe'' & (S/pixel) & \eqref{eq:zeta:soe} \\
			Percentage Average Deviation of Total Field Magnitude & $\zeta_{TFMPAD}$ & ``zeta\_tfmpad'' & (\%/pixel) & \eqref{eq:zeta:tfmpad} \\
			Average Deviation of Total Field Phase & $\zeta_{TFPAD}$ & ``zeta\_tfpad'' & (rad/pixel) & \eqref{eq:zeta:tfpad} \\
			Total Variation & $\zeta_{TV}$ & ``zeta\_tv'' & & \eqref{eq:zeta:tv} \\
			Position error & $\zeta_{P}$ & ``zeta\_p'' & (\%) & Alg.\ref{alg:zeta:p} \\
			Shape error & $\zeta_{S}$ & ``zeta\_s'' & (\%) & \footnotemark[\value{footnote}] \\
			\hline
		\end{tabular}
		\label{tab:result}
	\end{table*}
	
	\begin{algorithm}[!t]
		\caption{Proposed calculation for measuring the scatterer location error.}
		\begin{algorithmic}[1]
			\Procedure{Compute $\zeta_P$}{$\chi^g$, $\chi^r$}
			\State $TH\leftarrow \min(|\chi^r| + 0.5(\max(|\chi^r|)-\min(|\chi^r|)))$
			\State $\mathbf{M^g} \leftarrow \{ (i,j) ~|~ \chi^g_{ij} > 0, ~ \forall i = 1, \cdots, N_x,~ j = 1,\cdots,N_y \}$
			\State $\mathbf{M^r} \leftarrow \{ (i,j) ~|~ \chi^r_{ij} > TH, ~ \forall i = 1, \cdots, N_x,~ j = 1,\cdots,N_y \}$
			\State $x^g_c \leftarrow \frac{1}{|\mathbf{M^g}|} \sum\limits_{(i,j)\in M^g}\frac{x_{ij}+L_x/2}{L_x}$
			\State $y^g_c \leftarrow \frac{1}{|\mathbf{M^g}|} \sum\limits_{(i,j)\in M^g}\frac{y_{ij}+L_y/2}{L_y}$
			\State $x^r_c \leftarrow \frac{1}{|\mathbf{M^r}|} \sum\limits_{(i,j)\in M^r}\frac{x_{ij}+L_x/2}{L_x}$
			\State $y^r_c \leftarrow \frac{1}{|\mathbf{M^r}|} \sum\limits_{(i,j)\in M^r}\frac{y_{ij}+L_y/2}{L_y}$
			\State $\zeta_{P} \leftarrow 100\times\sqrt{(x^g_c-x^r_c)^2+(y^g_c-y^r_c)^2}$ (\%)
			\EndProcedure
		\end{algorithmic}
		\label{alg:zeta:p}
	\end{algorithm}
	
	This set of indicators can measure the performance of the algorithms, thus indicating their quality in locating, recovering the shape, estimating the electrical properties of the scatterers, estimating the total field, and quantifying the error in the equations. Although indicators for the reconstruction of tumor geometry have been recently proposed in the literature \cite{kurrant2021evaluating}, which are a kind of scatterer, our geometry and positioning error measurement methodology have been proposed to be used in general applications. Furthermore, in \cite{kurrant2021evaluating}, the geometry indicators depend on an image segmentation procedure using Machine Learning whereas, in our work, the calculation of $\zeta_{P}$ and $\zeta_{S}$ depends on simpler procedures. Therefore, $\zeta_{P}$ and $\zeta_{S}$ are original indicators in their formulation and contribute to the literature as a way to measure the quality of location and shape recovery of scatterers in a general EISP approach and using simpler operations.
	
	\subsection{Method Classes}\label{sub:method}
	
	Although our library was designed to address algorithms for the inverse problem, a framework for forward solvers is also available. These solvers can be used within the inverse methods or to synthesize test input data. Furthermore, given the need to execute methods with the same input and output pattern and the possibility of classifying the inverse solvers, abstract classes are used to represent the basic scope of the forward and inverse methods. Thus, using the inheritance mechanism, methods are implemented as derivations of these prototypes by overloading the necessary functions. In this way, the execution of the methods can be done indiscriminately. In the following subsections, classes for each of these types of methods will be introduced.
	
	\subsubsection{\textit{ForwardSolver}}\label{subsub:forward}
	
	The \textit{ForwardSolver} abstract class was implemented to represent the methods that determine the electric field based on the images of the electrical properties in $D$. In addition to calculating the scattered field in $S$, any implementation should be able to calculate the total field in $D$, and also, the incident one. The two main functions that need to be overloaded are ``solve'' and ``incident\_field'' in which the former must calculate the scattered field in $S$ and the total one in $D$, and the latter, the incident one in $D$. The incident field can be from either an infinite plane wave or any impressed source. However, the TMz polarization must be maintained. Until the date of this paper, only the implementation of the Method of Moments using the Conjugate-Gradient and Fast Fourier Transform (MoM-CG-FFT) \cite{su1987calculation,chen2018computational} is available.
	
	\subsubsection{\textit{InverseSolver}}\label{subsub:inverse}
	
\begin{figure}
		\centering
		\includegraphics[width=3.5in]{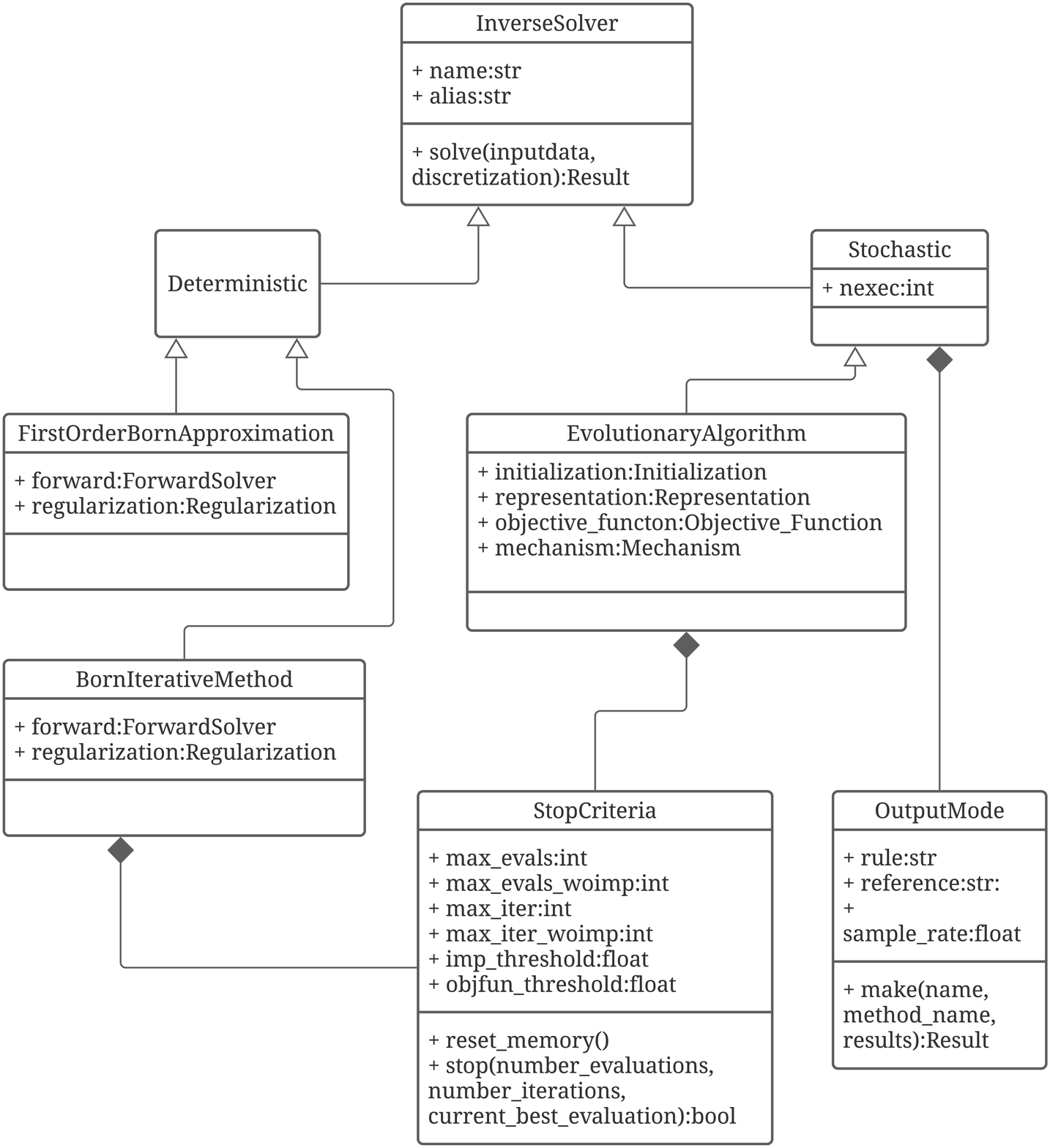}
		\caption{\textit{InverseSolver} UML Diagram Class.}
		\label{fig:uml:inversesolver}
	\end{figure}

	The abstract class \textit{InverseSolver} was implemented to represent the methods that determine the electrical properties and the total field in $D$. These will be divided into deterministic and stochastic methods, represented by the derived classes of the same name. This kind of differentiation was chosen as these characteristics are very important when an experiment is planned. Therefore, any method implementation must be a derivation of either the \textit{Deterministic} class or the \textit{Stochastic} class. Fig.\ref{fig:uml:inversesolver}  shows the UML Class Diagram for \textit{InverseSolver}.
	
	The function that must be overloaded in any implementation of the \textit{InverseSolver} class is ``run''. This function receives the input data, the form of discretization of the equations (Subsubsection \ref{subsub:discretization}), and executes the method. At the end of execution, the function must return an object of the \textit{Result} class containing information about the requested indicators and the reconstructed images.
	
	Any implementation of \textit{Stochastic} has the attribute ``nexec'', which means the number of executions that the algorithm should be repeated for any test. It means that a set of results is obtained at the end of the algorithm execution. However, in some situations, it is necessary to process these results to produce a single one that represents the performance of the algorithm. Due to this need, the \textit{OutputMode} class was implemented. This class aims to receive this list of results, process it, and define the overall performance of the algorithm based on a user-defined rule. Four rules are available:
	\begin{itemize}
		\item Best case scenario: the best result from the list is returned based on some user-defined indicator.
		\item Worst case: the worst result from the list is returned based on some user-defined indicator.
		\item Average case: the average of each indicator is returned. However, for cases where the indicator is calculated by iteration, a list is returned where each element is the average value for a given percentage of the iterations. The sampling rate in the range from 0 to 100\% can be controlled by the user. In addition, the images of the electrical and field properties of the solution, whose final value of the reference indicator is closest to the average, are stored in the final solution.
		\item Save each run: no processing is performed, and all results are returned.
	\end{itemize}
	
	Until the realise of this article, the following methods were available: BIM, First Order Born Approximation, and Evolutionary Algorithm (EA). These are implemented in the following classes, respectively: \textit{BornIterativeMethod}, \textit{FirstOrderBornApproximation}, and \textit{EvolutionaryAlgorithm}. The first two are derivations of the \textit{Deterministic} class and have as attributes a forward solver and a regularizer for the linear inverse problem (see Subsubsection \ref{subsub:regularization}). In the case of the Born Approximation, the forward resolver is only intended to provide the incident field. \textit{EvolutionaryAlgorithm}, on the other hand, is a class that can implement several forms of EAs, which will be presented in Subsubsection \ref{subsub:evoalglib}. Both \textit{EvolutionaryAlgorithm} and \textit{BornIterativeMethod} have an object of class \textit{StopCriteria} (Subsubsection \ref{subsub:stopcriteria}) to control the execution of algorithms, determining an end when some criterion is reached.
	
	\subsection{Experiments Classes}\label{sub:experiment}
	
	To structure the experimentation process, an abstract class, called \textit{Experiment}, was defined with two main attributes: a list of methods and a list of results. From this class, two classes are derived that implement a type of experimentation: \textit{CaseStudy} and \textit{Benchmark}.
	
	\subsubsection{\textit{CaseStudy}}\label{subsub:casestudy}
	
	The \textit{CaseStudy} class implements the study of the performance of algorithms on a specific test. This type of study is very useful for a preliminary assessment and to demonstrate the reconstructive capacity of a method. This class then has an attribute called ``test'', which is an object of the \textit{InputData} class. The class then allows executing a set of methods and analyzing the results graphically and statistically. In the case of stochastic methods, the class allows defining whether all results of all executions will be saved or not. However, some forms of result analysis are only possible if all results are stored. The following functions implement the following ways of analyzing the results:
	
	\begin{itemize}
		\item ``reconstruction'': shows the images reconstructed by the algorithms. They can be the images of the contrast function, the electrical properties, or the total field.
		\item ``convergence'': displays on a graph the convergence curve of an indicator for the different methods.
		\item ``boxplot'': graphically compares the values obtained by the stochastic algorithms in all executions considering a given indicator. The box plot informs the quantiles of 25, 50, and 75\% and the median of the values.
		\item ``compare'': statistically compares the averages of the results obtained by the stochastic algorithms for a given indicator. It can be used to compare a stochastic algorithm with a deterministic one or stochastic pairs or multiples.
		\item ``confint'': calculates the confidence interval for the average performance of stochastic algorithms on an indicator. The confidence level can be given by the user and the result can be shown graphically or textually.
	\end{itemize}
	
	\subsubsection{\textit{Benchmark}}\label{subsub:benchmark}
	
	The \textit{Benchmark} class implements the study of the average performance of algorithms in one or more test sets. By measuring performance through a set of tests, the effects of some factors can be eliminated (e.g., the geometry of scatterers), more robustly detect differences in method performance against some indicator, or more robustly verify if an algorithm meets some specific level of performance. Therefore, this kind of study can yield stronger conclusions about the impact of different choices in algorithms and the effect that different levels of some factors (e.g., noise) can have when multiple test sets are considered.
	
	This class has as attribute a single or a list of \textit{TestSet} objects (Subsubsection \ref{subsub:testset}). Regarding stochastic methods, all of them are configured to return the average case considering all executions performed for a single test. It is also possible to do the same analysis as the ``reconstruction'', ``boxplot'', ``compare'' and ``confint'' routines of the \textit{CaseStudy} class. However, for these last three tools, it is worth noting that each value in the graphs or samples is the result of an indicator for a test within the considered set. Therefore, these tools can be used for both deterministic and stochastic algorithms in this class. In addition to them, the following results analysis functions were implemented:
	
	\begin{itemize}
		\item ``plot'': displays in a graph the values obtained in each one of the tests of a set considering some indicator.
		\item ``violinplot'': displays on a graph the probability density regarding the indicator performance for a given method in a given test set.
		\item ``evolution'': when multiple test sets are considered, this routine displays the results (in boxplot format) of an indicator obtained in each test set in the desired sequence plus the line resulting from the linear regression of the means. The objective is to graphically analyze if the variation of any parameter in the configuration of the test sets causes an impact on the performance of the algorithm. It is also possible to use the same routine to analyze the evolution of performance when a parameter of the same algorithm varies.
		\item ``normality'': graphically compares the distribution of an indicator in a test set with the standard normal distribution. This feature aims to aid statistical comparisons.
	\end{itemize}
	
	\subsection{Auxiliary Classes and Modules}\label{sub:auxiliary}
	
	Some other classes and modules with specific functions have been implemented and are described in this subsection.
	
	\subsubsection{\textit{Discretization} Class}\label{subsub:discretization}
	
	An important choice for the solution of an EISP is the form of discretization of the equations. In addition to enabling the computational approach, it can be considered a way of regularizing the problem \cite{kirsch2021introduction}. The most used approach in methodologies for EISPs is Richmond's discretization \cite{richmond1965scattering}, given its simplicity of implementation, to represent the images and to analytically address the integral over the singularity in \eqref{eq:stateequation}. However, this approach is a particular case of the Collocation Method \cite{fletcher1984computational}. Another methodology for discretizing both differential and integral equations is the Galerkin Method \cite{kirsch2021introduction,fletcher1984computational}.
	
	Based on this hierarchy, an abstract class, called \textit{Discretization}, was implemented to represent the general scope of any discretization methodology. Any implementation must be able to calculate the residuals of the data \eqref{eq:dataequation} and states \eqref{eq:stateequation} equations; calculate the scattered field from the current solutions of $\chi$ and $E_z$ from \eqref{eq:dataequation}; solve systems of linear equations formed from the problem variables, and provide contrast function and full-field images as per the resolution required by the user. From the \textit{Discretization} class, the \textit{Collocation} class was derived, which represents a general scheme of the Collocation Method formulations. From the \textit{Collocation} class, the \textit{Richmond} class was derived that implements the discretization proposed in \cite{richmond1965scattering} with all the specifics of the general scope of the abstract class. The \textit{Richmond} class has as an attribute a tuple $(N_y, N_x)$ that represents the number of elements in the $y$- and $x$-axis of the image.
	
	\subsubsection{\textit{TestSet} Class}\label{subsub:testset}
	
	A test set is a sample of a universe of problems that share some common characteristics. In EISP, characteristics of scatterers such as contrast level, size, amount, and pattern of geometry can influence the performance of the algorithms. All of these factors are related to the DNL of each instance. However, the noise level is an impacting factor that is not associated with DNL.
	
	Based on this, the \textit{TestSet} class was implemented to generate and store a set of tests. For the generation of tests, the following control parameters were defined:
	
	\begin{itemize}
		\item Maximum contrast: maximum level of contrast allowed for the scatterers;
		\item Contrast mode: assign all scatterers with the same contrast level or allow them to vary up to the defined maximum value;
		\item Maximum size: maximum radius allowed between the center of the scatterer to its farthest boundary;
		\item Size mode: fix the same radius between the center and the furthest boundary of all scatterers or allow it to vary up to the maximum defined value;
		\item Density: number of scatterers (maximum or fixed) or maximum allowed value of the contrast (absolute value) average per pixel;
		\item Density mode: set the number of scatterers or the maximum allowed or control the average contrast per pixel.
		\item Map pattern: Fix the geometry of the scatterers on either regular polygons or random polygons or surfaces (Subsubsection \ref{subsub:draw}).
		\item Noise Level: Amount of noise added to the scattered field data. Noise is modeled as a phasor with random phase (uniform distribution) and user-controlled magnitude expressed in terms of the percentage of magnitude of the original value.
		\item Sample size: number of tests present in the set.
	\end{itemize}
	
	Therefore, the universe of problems is defined based on the configuration of these parameters according to the interest of the study. The generated tests are samples from this universe, and the scattered field data can be synthesized through a forward solver. Finally, it is important to emphasize that, in the case of the pattern of the maps configured for surfaces, the parameters of Maximum contrast, Maximum size, and Density control the height, width, and amount of elevations in the image.
	
	\subsubsection{\textit{StopCriteria} Class}\label{subsub:stopcriteria}
	
	Many methods rely on defining some stopping criteria to delimit their executions. With this in mind, the \textit{StopCriteria} class was implemented to aggregate a set of different stopping criteria that can be defined by users and can be included in methods to control the duration of their processes. The following stop criteria have been defined:
	
	\begin{itemize}
		\item Maximum number of iterations;
		\item Maximum number of iterations without improvement;
		\item Maximum number of evaluations;
		\item Maximum number of evaluations without improvement;
		\item Threshold of the objective function.
	\end{itemize}
	
	It is important to highlight that, in the context of EAs, an evaluation means a call for calculating the objective function of the problem. Stopping criteria based on evaluations are important as they allow the comparison of EA formulations that involve different amounts of solutions. Although this is not common in deterministic methods, these criteria can be used when it is necessary to count how many times a specific operation is performed, which is not necessarily called once per iteration.
	
	From the definition of these criteria in the construction of the object, the stop check can be done from the ``stop'' function that receives the current number of iterations, evaluations, and the value of the objective function and returns a Boolean variable indicating whether the algorithm should stop or not. A routine to reset the counts and history of the three input parameters was also implemented.
	
	\subsubsection{\textit{Regularization} Class}\label{subsub:regularization}
	
	Regularizers play an important role in ill-posed inverse problems \cite{kirsch2021introduction}. Therefore, they are used in many EISP methodologies that rely on some kind of problem linearization (e.g., BIM) \cite{chen2018computational}. Since there are many regularization strategies, an abstract class called \textit{Regularization} was created, which implements the basic scope of operation. All implementations solve the problem represented by the matrix $\mathbf{K}$ and column-vector $\mathbf{y}$ which stand for the kernel and the right-hand side of the equation, respectively. The methods implemented so far in this article are listed below, along with their respective parameters:
	
	\begin{itemize}
		\item Tikhonov: the regularization parameter can be user-defined or can be calculated based on either Morozov's Discrepancy Principle or L-Curve \cite{kirsch2021introduction};
		\item Landweber and Gradient-Conjugate: the regularization parameter is the number of iterations and must be defined in the input;
		\item Spectral Cut-Off: the regularization parameter is the cut-off threshold of the singular values;
	\end{itemize}
	
	\subsubsection{\statistics Module}\label{subsub:statistics}
	
	All necessary routines for statistical comparisons and confidence interval calculations of means are in the \statistics module. The module uses many functions already available in the Scipy \cite{scipy} and Statsmodels \cite{statsmodels} libraries. In addition, the Dunnett Test and the Factorial Analysis for two or three factors were also implemented \cite{montgomery2017design}.
	
	Based on the input arguments of the ``compare'' functions implemented for the \textit{CaseStudy} and \textit{Benchmark} classes, one of the comparison routines available in \statistics is executed: ``compare1sample'', ``compare2samples'', and ``compare\_multiple''. Algorithms \ref{alg:compare:1} and \ref{alg:compare:2} summarize each of first two routines, respectively, and inform which statistical tests are used in each case. Algorithms \ref{alg:compare:multiple:paired} and \ref{alg:compare:multiple:independent} summarize the ``compare\_multiple'' routine when the paired design is required or not, respectively. Multiple comparisons of paired samples are required when one wants to analyze the results of multiple methods for a single test set (the results of each test among the algorithms must be paired). Alternatively, when analyzing multiple executions of the same test or different test sets, the samples should be treated independently.
	
	The following observations are necessary: (i) the verification of the normality assumption is made by comparing the p-value of the Shapiro-Wilk Test on the involved sample with a significance level of 0.05; (ii) the verification of the homoscedasticity assumption is made by comparing the p-value of the Fligner-Killeen Test on the samples involved with a significance level of 0.05; (iii) we chose not to forego parametric tests when possible to minimize the false-negative error in comparisons; (iv) in some cases the routines also return the effect size to 80\% power.
	
	\begin{algorithm}[!t]
		\centering
		\caption{Outline of the implemented procedure for testing the mean of a single sample $\mathbf{x}$ against the expected value $\mu_{H0}$ .}
		\begin{algorithmic}[1]
			\Procedure{Compare Single Sample}{$\mathbf{x}$, $\mu_{H0}$}
			\If{normality assumption holds}
			\State Perform the Student's T-Test to test $\mu_\mathbf{x}=\mu_{H0}$ against the two-sided alternative hypothesis and, if necessary, against the greater or less ones.
			\Else
			\State Perform the Wilcoxon signed-rank test to test if $\mathbf{x}-\mu_{H0}$ is symmetric about zero and, if necessary, against the greater or less alternative hypothesis.
			\EndIf
			\EndProcedure
		\end{algorithmic}
		\label{alg:compare:1}
	\end{algorithm}
	
	\begin{algorithm}[!t]
		\centering
		\caption{Outline of the implemented procedure for testing the paired or unpaired means of two samples $\mathbf{x_1}$ and $\mathbf{x_2}$.}
		\begin{algorithmic}
			\Procedure{Compare Two Samples}{$\mathbf{x_1}$, $\mathbf{x_2}$}
			\If{Paired Design}
			\If{normality assumption holds}
			\State Perform the Paired Student's T-Test to test $\mathbf{x_1} = \mathbf{x_2}$ against the two-sided alternative hypothesis and, if necessary, against the greater or less ones.
			\Else
			\State Perform the Wilcoxon signed-rank test to test if $\mathbf{x_1}-\mathbf{x_2}$ is symmetric about zero and, if necessary, against the greater or less alternative hypothesis.
			\EndIf
			\Else
			\If{normality assumption holds}
			\If{homocedascity assumption holds}
			\State Perform the Two Sample Student's T-Test to test $\mathbf{x_1} = \mathbf{x_2}$ against the two-sided alternative hypothesis and, if necessary, against the greater or less ones
			\Else
			\State Perform the Welch's T-Test to test $\mathbf{x_1} = \mathbf{x_2}$ against the two-sided alternative hypothesis and, if necessary, against the greater or less ones
			\EndIf
			\Else
			\State Perform the Mann-Whitney U rank test to test if the distribution underlying $\mathbf{x_1}$ is the same as the distribution underlying sample $\mathbf{x_2}$ and, if necessary, against the greater or less alternative hypothesis.
			\EndIf
			\EndIf
			\EndProcedure
		\end{algorithmic}
		\label{alg:compare:2}
	\end{algorithm}
	
	\begin{algorithm}[!t]
		\centering
		\caption{Outline of the implemented procedure for testing multiple samples according paired design}
		\begin{algorithmic}[1]
			\Procedure{Compare Multiple Paired Samples}{$\mathbf{x_1}$, $\mathbf{x_2}$, $\cdots$, $\mathbf{x_N}$}
			\If{normality assumption of residuals holds}
			\State Perform Randomized Complete Block Design to test if the pairwise difference among $\mathbf{x_1}$, $\mathbf{x_2}$, $\cdots$, $\mathbf{x_N}$ do not deviate from zero.
			\If{means are not equal and pairwise comparisons are required}
			\State Perform multiple Paired T-Test with Bonferroni correction of the significance level
			\ElsIf{means are not equal and all vs. one study is required}
			\State Perform multiple Paired T-Test with Bonferroni correction of the significance level
			\EndIf
			\Else
			\State Perform Friedman's test to verify if the pairwise difference among $\mathbf{x_1}$, $\mathbf{x_2}$, $\cdots$, $\mathbf{x_N}$ do not deviate from zero.
			\If{medians are not equal and pairwise comparisons are required}
			\State Perform the Wilcoxon's test to test if the distribution underlying each pair of samples is the same.
			\ElsIf{means are not equal and all vs. one study is required}
			\State Perform the Wilcoxon's test to test if the distribution underlying each sample is the same as the control one.
			\EndIf
			\EndIf
			\EndProcedure
		\end{algorithmic}
		\label{alg:compare:multiple:paired}
	\end{algorithm}
	
	\begin{algorithm}[!t]
		\centering
		\caption{Outline of the implemented procedure for testing multiple independent samples}
		\begin{algorithmic}[1]
			\Procedure{Compare Multiple Independent Samples}{$\mathbf{x_1}$, $\mathbf{x_2}$, $\cdots$, $\mathbf{x_N}$}
			\If{normality assumption of residuals holds}
			\If{homocedascity assumption of residuals holds}
			\State Perform One-Way ANOVA to test if $\mathbf{x_1}$, $\mathbf{x_2}$, $\cdots$, $\mathbf{x_N}$ have the same mean values.
			\If{means are not equal and pairwise comparisons are required}
			\State Perform Tukey's HSD test
			\ElsIf{means are not equal and all vs. one study is required}
			\State Perform Dunnett's test
			\EndIf
			\Else
			\State Perform Welch's ANOVA to test if $\mathbf{x_1}$, $\mathbf{x_2}$, $\cdots$, $\mathbf{x_N}$ have the same mean values.
			\If{means are not equal and pairwise comparisons are required}
			\State Perform multiple two sample Welch's T-Test with Bonferroni correction of the significance level
			\ElsIf{means are not equal and all vs. one study is required}
			\State Perform multiple two sample Welch's T-Test with Bonferroni correction of the significance level
			\EndIf
			\EndIf
			\Else
			\State Perform the Kruskal-Wallis H-test to test if the population median of all of the groups are equal.
			\If{medians are not equal and pairwise comparisons are required}
			\State Perform the Mann-Whitney U rank test to test if the distribution underlying each pair of samples is the same.
			\ElsIf{means are not equal and all vs. one study is required}
			\State Perform the Mann-Whitney U rank test to test if the distribution underlying each sample is the same as the control one.
			\EndIf
			\EndIf
			\EndProcedure
		\end{algorithmic}
		\label{alg:compare:multiple:independent}
	\end{algorithm}
	
	\subsubsection{\evoalglib Module}\label{subsub:evoalglib}
	
	Evolutionary algorithms are a class of stochastic methods inspired by biological or physical processes \cite{eiben2003introduction}. Although there are many different formulations, EAs are usually composed of a set of candidate solutions, called population, which is iteratively updated to explore new solutions until a stopping criterion is reached, and the best solution found concerning the optimization of the objective function is returned. To enable the numerous possibilities of formulations within the context of EISPs, the structure of EAs is divided into four fundamental components: the representation of solutions, objective function, population initialization methodology, and the evolutionary mechanism. For each of these components, an abstract class was implemented to define the general operating scope. It is important to highlight that the \textit{EvolutionaryAlgorithm} class (Subsubsection \ref{subsub:inverse}) requires an object of each of these classes. The abstract classes corresponding to each of these components are summarized with the derivations available so far:
	
	\begin{itemize}
		\item \textit{Representation}: this class implements the ways of translating the EISP unknowns into a set of EA decision variables. An example in the literature is the representation of the contrast function through the scatterers contours, mathematically described by a series of functions \cite{chiu1996image,qing2006dynamic,huang2008time}. In this way, EA decision variables are the coefficients of the series. Another alternative is to interpret the value of each element of the discretization of equations as a decision variable \cite{salucci2017multifrequency,donelli2010differential,yang2021quantum}. In this way, when Richmond's discretization is used, the contrast value in each pixel of the image is a variable to be determined by EA. The disadvantage is that the number of variables can become very large, especially when the total field is also determined in the same fashion. However, this approach does not depend on assumptions on geometry and the number of scatterers in the image. Currently, only this last type of representation is implemented through class \textit{DiscretizationElementBased}.
		\item \textit{Objective-function}: the objective-function formulations are implemented in this class. In literature, the most traditional one is the weighted sum of residuals from data and state equations \cite{pastorino2010ch6} and is available through the \textit{WeightedSum} class.
		\item \textit{Initialization}: in this class, the ways to initialize the values of the decision variables of each solution are implemented. A simple methodology is to initialize the population with random values based on a uniform distribution. This methodology is implemented in the \textit{UniformRandomDistribution} class. Another approach is to initialize only the variables corresponding to the contrast function with uniformly distributed random values and calculate the total field based on First-Order Born Approximation \cite{salucci2017multifrequency}. This methodology assumes the contrast and field solution from the representation based on discretization elements and is implemented in the \textit{BornApproximation} class.
		\item \textit{Mechanism}: in this class, all the evolutionary search mechanisms are implemented. The following mechanisms are currently available: Particle Swarm Optimization (PSO), Differential Evolution (DE), and Genetic Algorithm (GA). Each of these mechanisms is implemented in classes \textit{ParticleSwarmOptimization}, \textit{DifferentialEvolution}, and \textit{GeneticAlgorithm}, respectively.
	\end{itemize}
	
\begin{figure}
		\centering
		\includegraphics[width=3.5in]{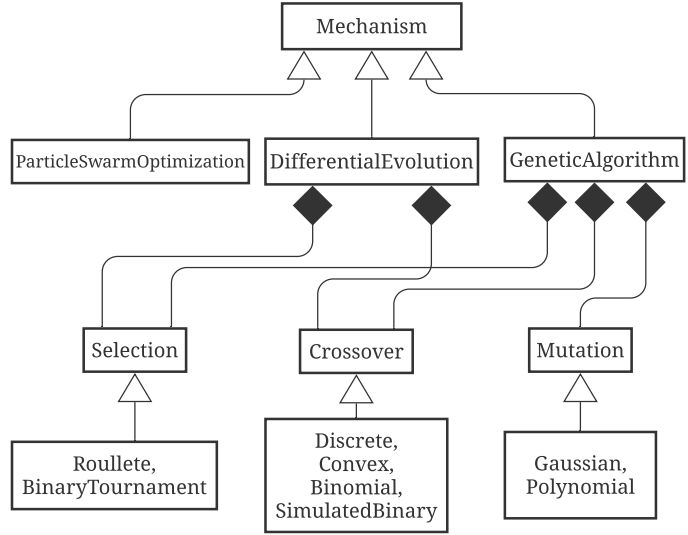}
		\caption{Mechanism UML Class Diagram.}
		\label{fig:uml:mechanism}
	\end{figure}

	It is worth noting that these evolutionary mechanisms can be formulated with different crossover, mutation, and selection operators \cite{eiben2003introduction}. Therefore, different versions of these operators have been implemented, as can be viewed in Fig.\ref{fig:uml:mechanism}.
	
	In order to enable the testing of EA formulations in canonical optimization problems, classes \textit{Rastringin}, \textit{Rosenbrock}, and \textit{Ackley}, derived from class \textit{ObjectiveFunction}, were implemented. These are traditional objective functions in the context of non-linear optimization \cite{li2013benchmark} and have similar characteristics to EISPs such as non-linearity, multi-modality, and smooth local irregularities. To enable the representation of unconstrained optimization problems with continuous real variables, the class \textit{CanonicalProblems}, derived from \textit{Representation}, was implemented. Therefore, these tools can support the preliminary study of the application of new ideas in EAs to EISPs.
	
	Finally, all implementations mentioned in this subsubsection are gathered in module \evoalglib.
	
	\subsubsection{\draw Module}\label{subsub:draw}
	
	An important aspect of testing algorithms for EISP is the geometry of scatterers. Recovering different shapes can be an attractive characteristic for an algorithm besides being essential in some applications. Therefore, the \draw module has been developed, which implements a set of routines for drawing geometric figures in images. These routines are then designed to be used in the creation of tests, specifically in defining the images of relative permittivity and conductivity. It is possible to draw: 
	
	\begin{itemize}
		\item Traditional polygons: square, equilateral triangle, star (4, 5, 6 points), ring, circle, ellipse, cross, line, rhombus, trapezoid, parallelogram, pentagon, hexagon, heptagon, and octagon.
		\item Random polygons: from the definition of the number of sides, a polygon is generated whose distance between the vertices and the center is defined randomly.
		\item Surfaces: formed by the sum of cosine functions or Gaussians curves.
	\end{itemize}
	
	Besides the possibility of defining the images in any resolution, the polygons can be positioned and rotated according to the user specification.
	
	\section{Usage Examples}\label{sec:example}
	
	In this section, three examples are presented to illustrate common situations of \textbf{eispy2d} library use\footnote{Script and notebook versions of all examples are available at \url{https://andre-batista.github.io/eispy2d/usage-examples.html}}. The first consists of a manual definition of a test, its solution through a method, and viewing its results. In the second example, the use of the \textit{CaseStudy} class is illustrated. Finally, an example of the use of class \textit{Benchmark} is presented.
	
	\subsection{Manual Definition of a Test}\label{sub:manual}
	
	In this example, BIM will be tested on an instance characterized by a 5-pointed star positioned in the lower right corner of the image. First, the domains will be specified along with the characteristics of the source and the background medium by defining an object of class \textit{Configuration}. The incidence of 25 plane waves ($N_S = 25$) propagating in a vacuum ($\epsilon_{rb} = 1$, $\sigma_b = 0$ [S/m]) is considered, with a magnitude ($E_0$) of 1 (V/m) and wavelength ($\lambda_b$) equal to 1 (m). Domain $D$ will be defined as a square ($L_x=L_y$) whose length is equal to 4$\lambda_b$. The radius ($R_O$) of Domain $S$ will be defined as 6$\lambda_b$, and the scattered field will be sampled at 25 points ($N_M=25$). The problem assumes only perfectly dielectric scatterers. The code below implements this specification:
	
	{\footnotesize 
\begin{lstlisting}[language=Python]
import configuration as cfg
config = cfg.Configuration(
    name='cfg_test',
    wavelength=1.,
    wavelength_unit=True,
    number_measurements=25,
    number_sources=25,
    image_size=[4., 4.],
    observation_radius=6.,
    background_permittivity=1.,
    magnitude=1.,
    perfect_dielectric=True
)
\end{lstlisting}}
	
	Next, the input data is built. To accomplish this, an object of class \textit{InputData} is initialized. The built configuration is assigned to the object and specify, first, the resolution of the test images, the noise level, and the indicators that will be adopted. The test image resolution will be set to 100$\times$100 pixels. The noise level will be specified as 1 (\%/sample). The indicators that will be adopted in this example are: $\zeta_{RN}$, $\zeta_{RPAD}$, $\zeta_{\epsilon PAD}$, $\zeta_{\epsilon OE}$, $\zeta_{\epsilon BE}$, $\zeta_{P}$, and $\zeta_{S}$. The total execution time is also stored. The following code implements this specification:
	
	{\footnotesize 
\begin{lstlisting}[language=Python]
import inputdata as ipt
import result as rst
inputdata = ipt.InputData(
    name='ipt_test',
    configuration=config,
    resolution=(100, 100),
    noise=1.,
    indicators=[
        rst.RESIDUAL_PAD_ERROR,
        rst.RESIDUAL_NORM_ERROR,
        rst.REL_PERMITTIVITY_PAD_ERROR,
        rst.REL_PERMITTIVITY_OBJECT_ERROR,
        rst.REL_PERMITTIVITY_BACKGROUND_ERROR,
        rst.POSITION_ERROR,
        rst.SHAPE_ERROR,
        rst.EXECUTION_TIME
    ]
)
\end{lstlisting}}
	
	The next step is to create the scatterer image. For this, 5-pointed star will be considered through the appropriated \draw routine. The radius between the center and the points of the star will be fixed at $\lambda_b$. The relative permittivity of the scatterer will be set to 1.25 ($\chi = 0.25$). The coordinates ($x$, $y$) of the center of the star will be fixed at (0.8m, -0.8m). The star will also be rotated 25 degrees. The following code implements the scatterer definition:
	
	{\footnotesize 
\begin{lstlisting}[language=Python]
import draw
contrast = .25
inputdata.rel_permittivity, _ = draw.star5(
    1.,
    axis_length_x=config.Lx,
    axis_length_y=config.Ly,
    resolution=inputdata.resolution,
    background_rel_permittivity=config.epsilon_rb,
    object_rel_permittivity=(
        config.epsilon_rb + contrast
    ),
    center=[-.8, .8],
    rotate=25
)
inputdata.draw(fontsize=20, save=True)
inputdata.compute_dnl()
\end{lstlisting}}
	
\begin{figure}
		\centering
		\includegraphics[width=3.5in]{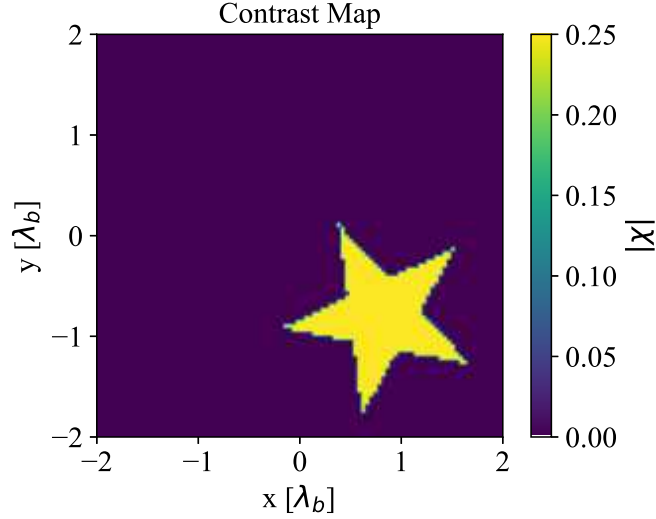}
		\caption{Image of the scatterer used in the manual definition example.}
		\label{fig:manual:scatterer}
	\end{figure}

	The problem image generated by the previous code is presented in Fig.\ref{fig:manual:scatterer}. The DNL of this instance is rated at 0.2719, which means that this is a weak scatterer, and the method should be able to recover it without much difficulty.
	
	Then it is necessary to synthesize the scattered field data. For this, the MoM-CG-FFT will be used with a tolerance of 0.001 and 5000 iterations maximum. The following code implements this task:
	
	{\footnotesize 
\begin{lstlisting}[language=Python]
import mom_cg_fft as mom
solver = mom.MoM_CG_FFT(tolerance=.001,
                        maximum_iterations=5000)
_ = solver.solve(inputdata,
                 COMPUTE_SCATTERED_FIELD=True,
                 SAVE_INTERN_FIELD=False)
\end{lstlisting}}
	
	Before defining the method, the discretization, in which the equations will be solved, needs to be defined. The Richmond's discretization is chosen and the number of elements is set to 50 in each direction. In this way, the inverse crime is avoided \cite{colton2019inverse,wirgin2004inverse} since the scattered field data was obtained using the original image resolution. The following code implements this choice:
	
	{\footnotesize 
\begin{lstlisting}[language=Python]
import richmond as ric
elements = (50, 50)
discretization = ric.Richmond(
    config,
    elements
)
\end{lstlisting}}
	
	Now we define the inverse solver. The BIM will be equipped with both MoM-CG-FFT to solve the forward linear problem and Tikhonov Regularization to solve the inverse linear subproblem. The version of MoM-CG-FFT used by the method is defined with a tolerance of 0.01 and 2500 iterations, maximum. The Tikhonov regularization parameter will be set to 0.01. The stopping criterion of the algorithm will be established in 5 iterations. The next code performs this implementation:
	
	{\footnotesize 
\begin{lstlisting}[language=Python]
import bim
import regularization as reg
import stopcriteria as stp
method = bim.BornIterativeMethod(
    mom.MoM_CG_FFT(
        tolerance=0.01,
        maximum_iterations=2500
    ),
    reg.Tikhonov(reg.TIK_FIXED,
                 parameter=0.1),
    stp.StopCriteria(
        max_iterations=5
    )
)
\end{lstlisting}}
	
	All the components required for running the algorithm and visualizing the results are now set. The commands for executing the method, viewing the recovered image, and the convergence of the indicators can be checked as follows:
	
	{\footnotesize 
\begin{lstlisting}[language=Python]
result = method.solve(inputdata, discretization)
result.plot_map(show=True, fontsize=15)
result.plot_convergence(
    indicators=rst.RESIDUAL_NORM_ERROR,
    show=True
)
result.plot_convergence(
    indicators=rst.RESIDUAL_PAD_ERROR,
    show=True
)
result.plot_convergence(
    indicators=rst.REL_PERMITTIVITY_PAD_ERROR,
    show=True
)
result.plot_convergence(
    indicators=rst.REL_PERMITTIVITY_OBJECT_ERROR,
    show=True
)
result.plot_convergence(
    indicators=rst.REL_PERMITTIVITY_BACKGROUND_ERROR,
    show=True)
result.plot_convergence(
    indicators=rst.SHAPE_ERROR,
    show=True
)
result.plot_convergence(
    indicators=rst.POSITION_ERROR,
show=True
)
\end{lstlisting}}
	
\begin{figure*}
		\centering
		\subfloat[]{\includegraphics[width=.25\textwidth]{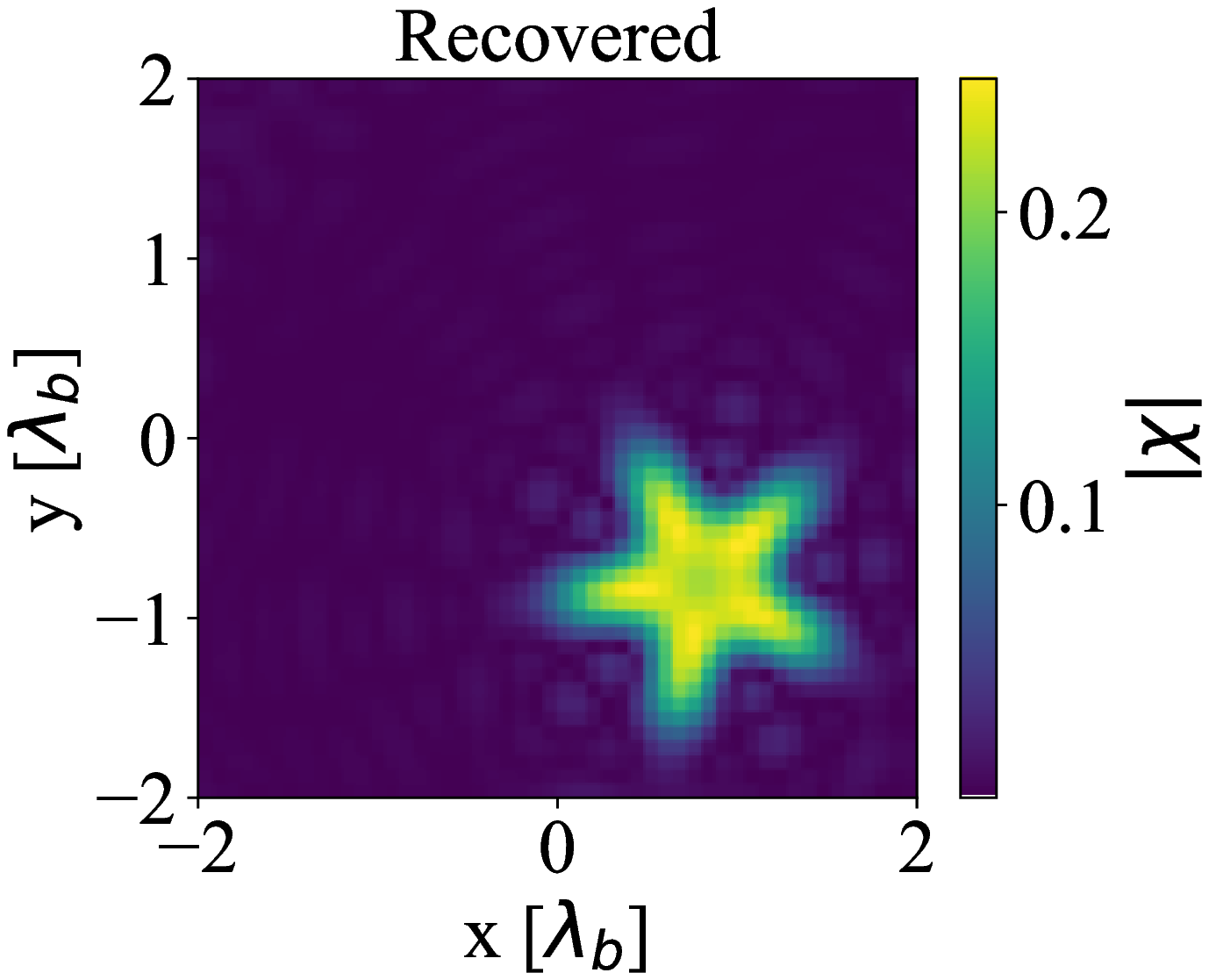}} 
		\subfloat[]{\includegraphics[width=.25\textwidth]{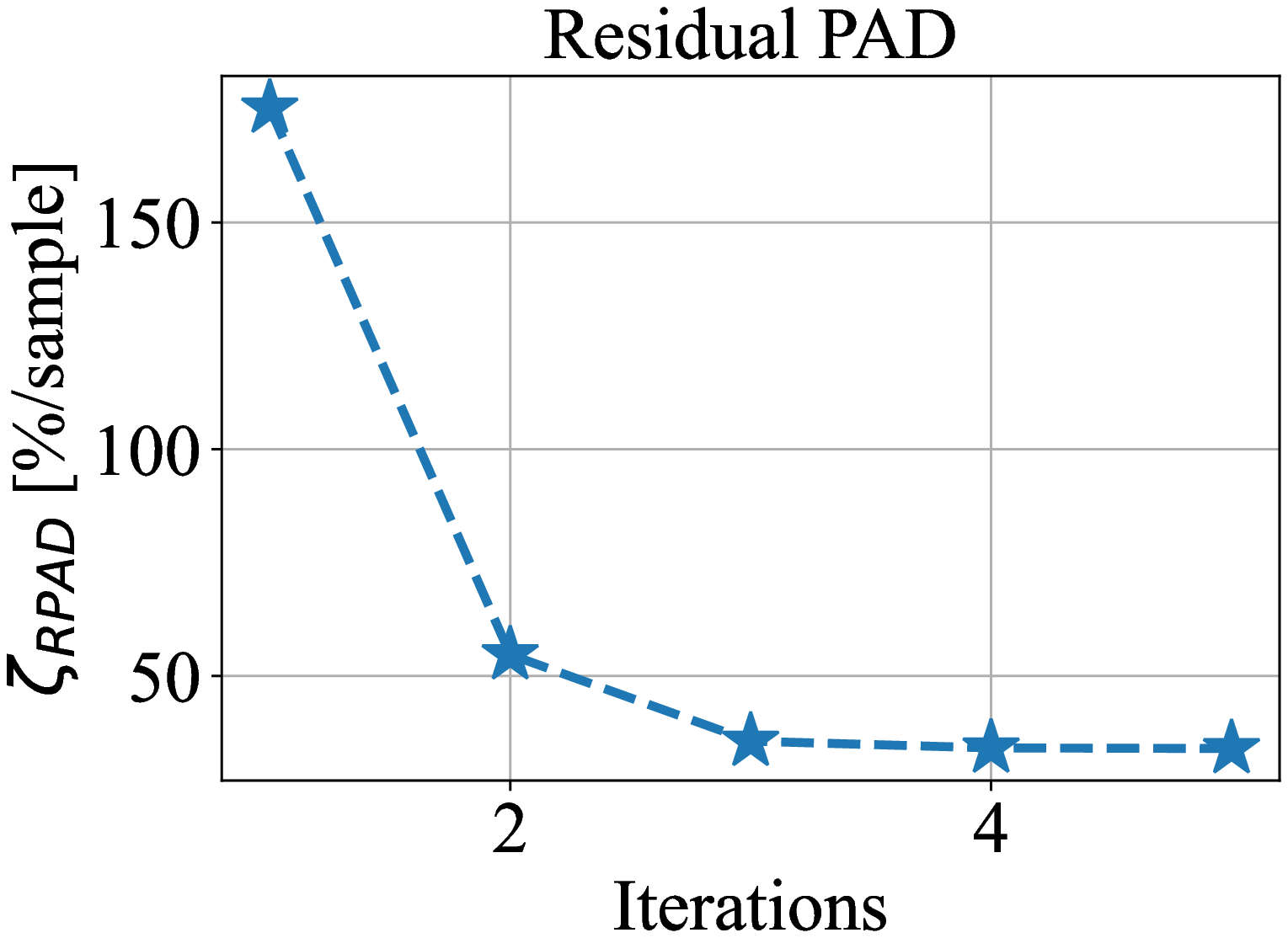}}
		\subfloat[]{\includegraphics[width=.25\textwidth]{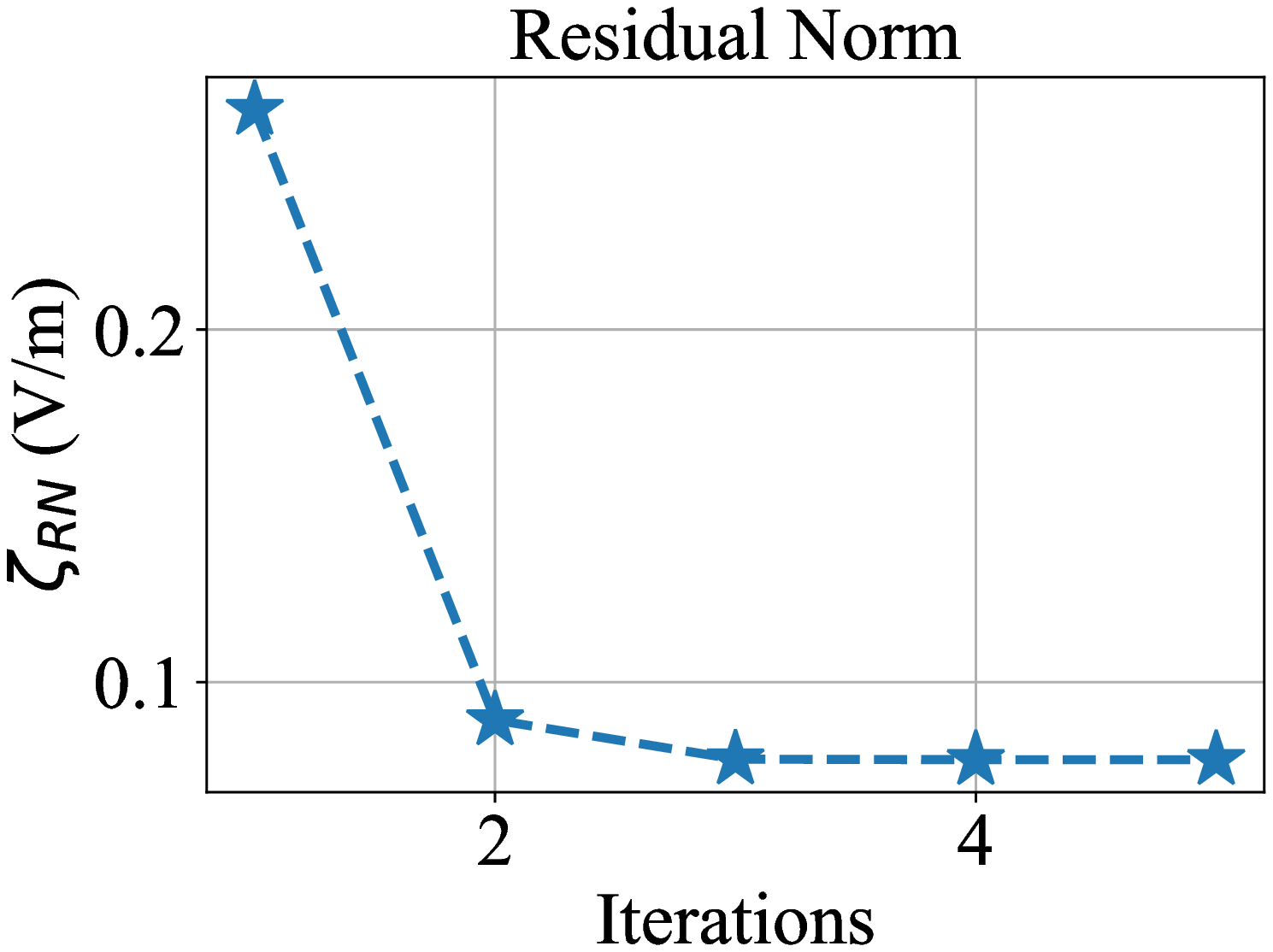}}
		\subfloat[]{\includegraphics[width=.25\textwidth]{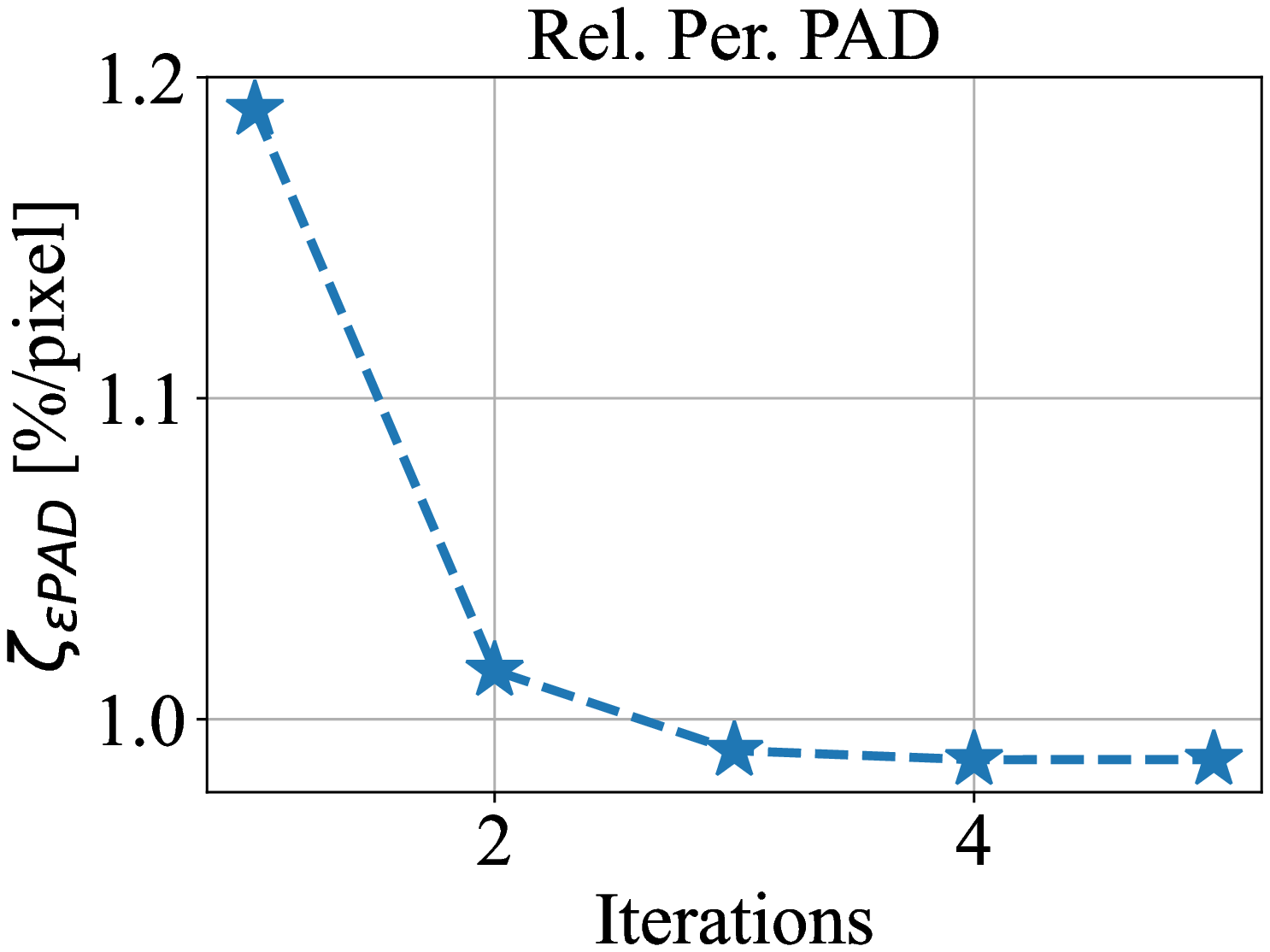}} \\
		\subfloat[]{\includegraphics[width=.25\textwidth]{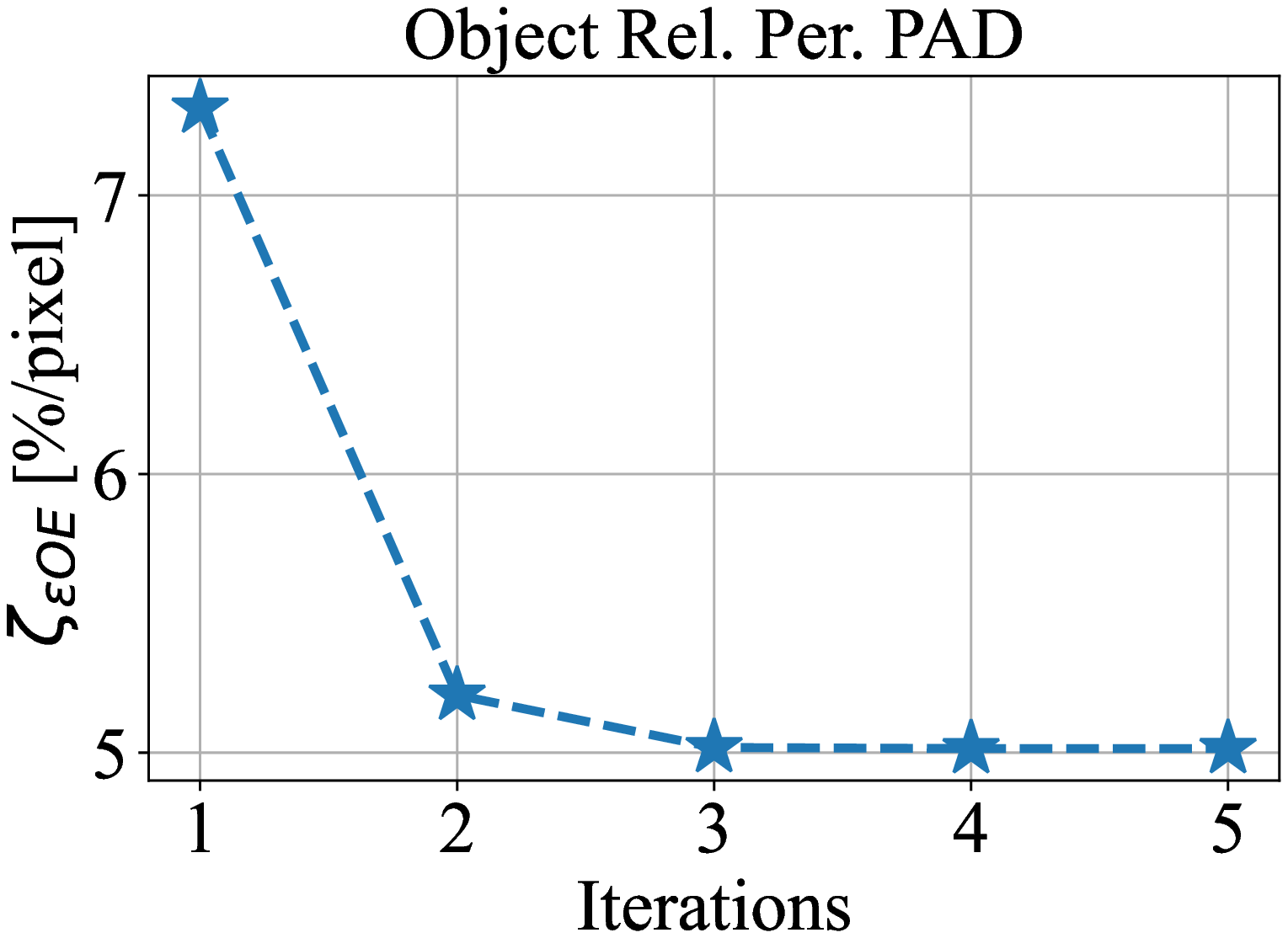}}
		\subfloat[]{\includegraphics[width=.25\textwidth]{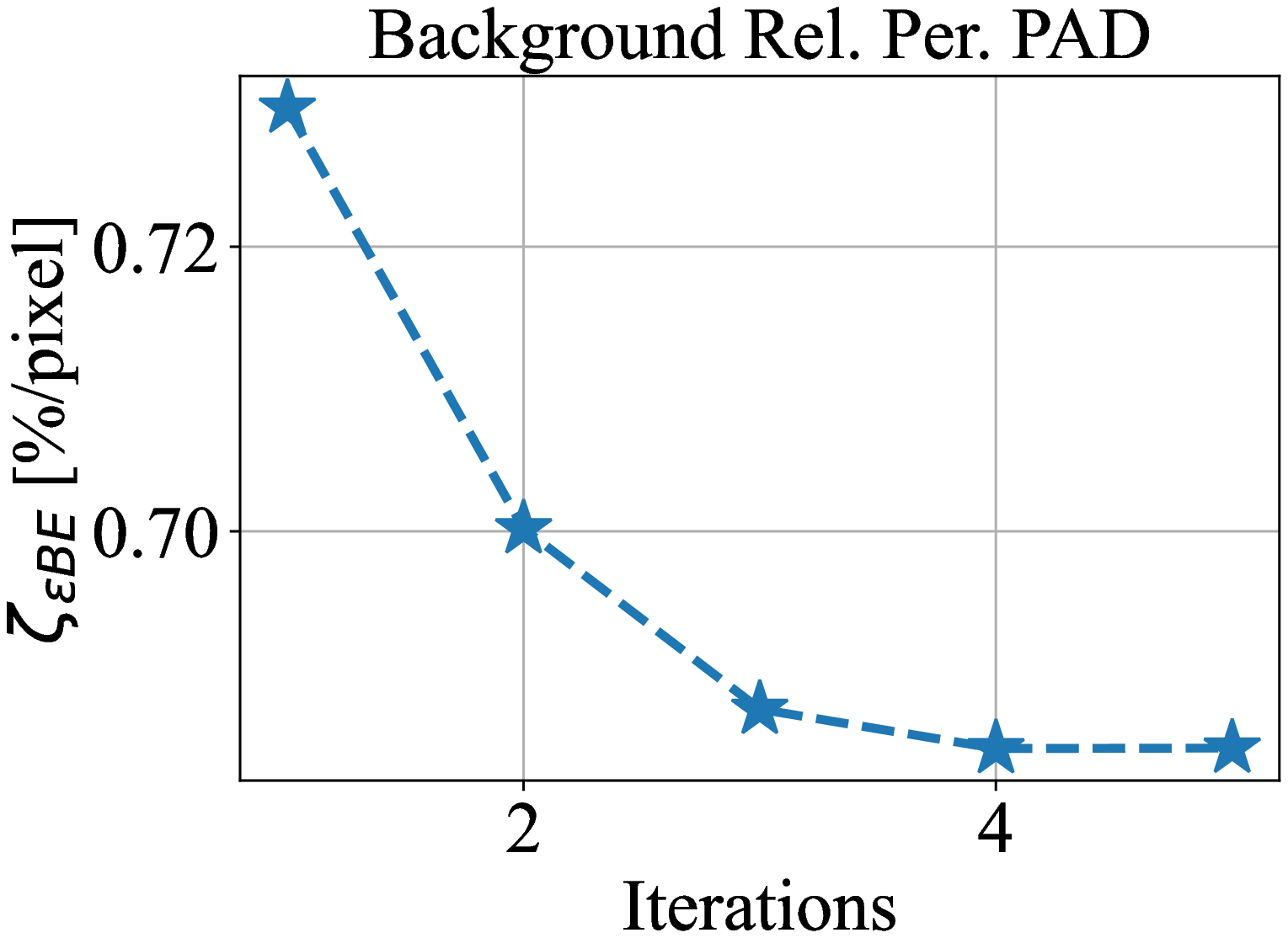}}
		\subfloat[]{\includegraphics[width=.25\textwidth]{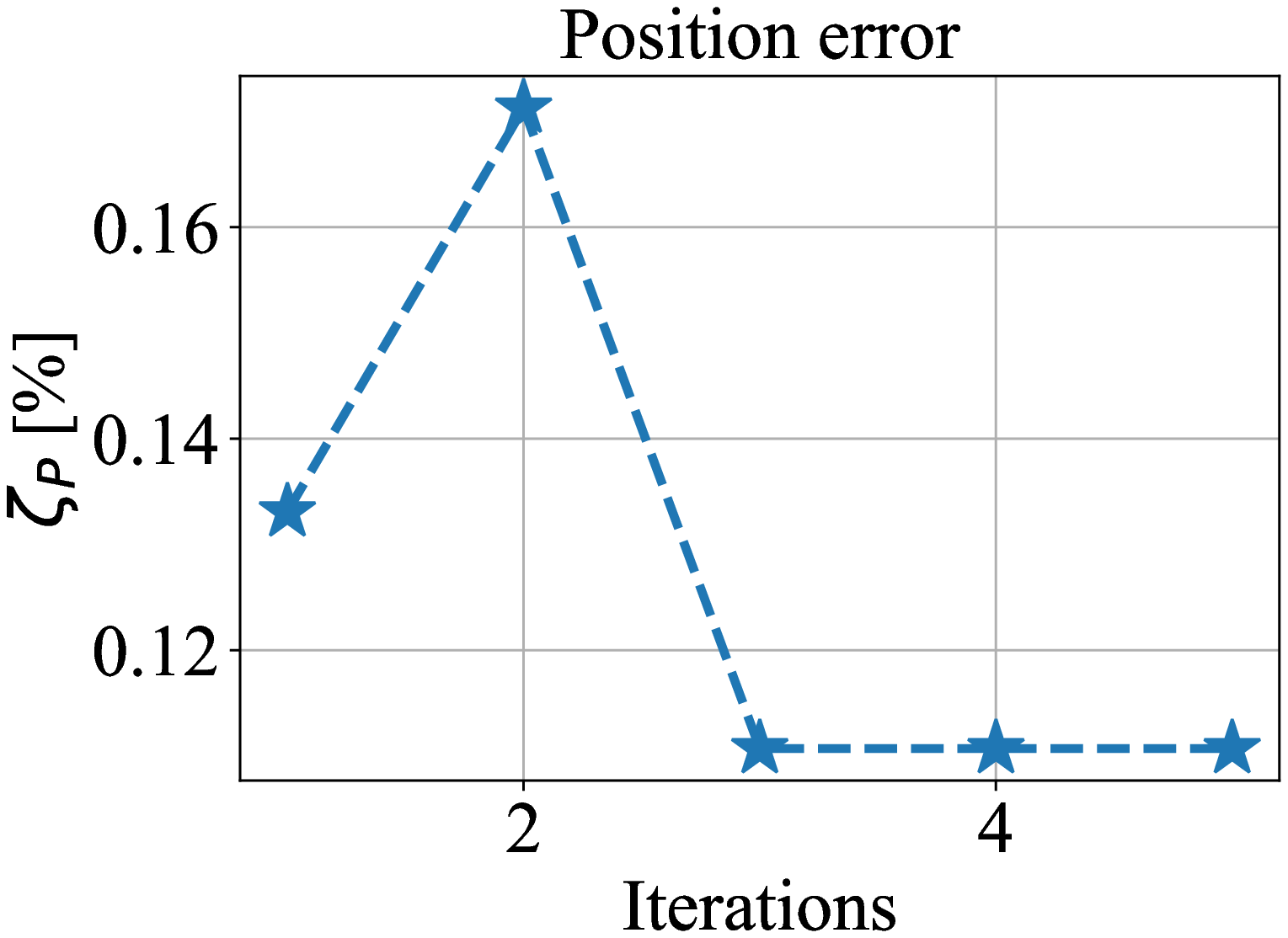}\label{fig:manual:results:zeta_s}}
		\subfloat[]{\includegraphics[width=.25\textwidth]{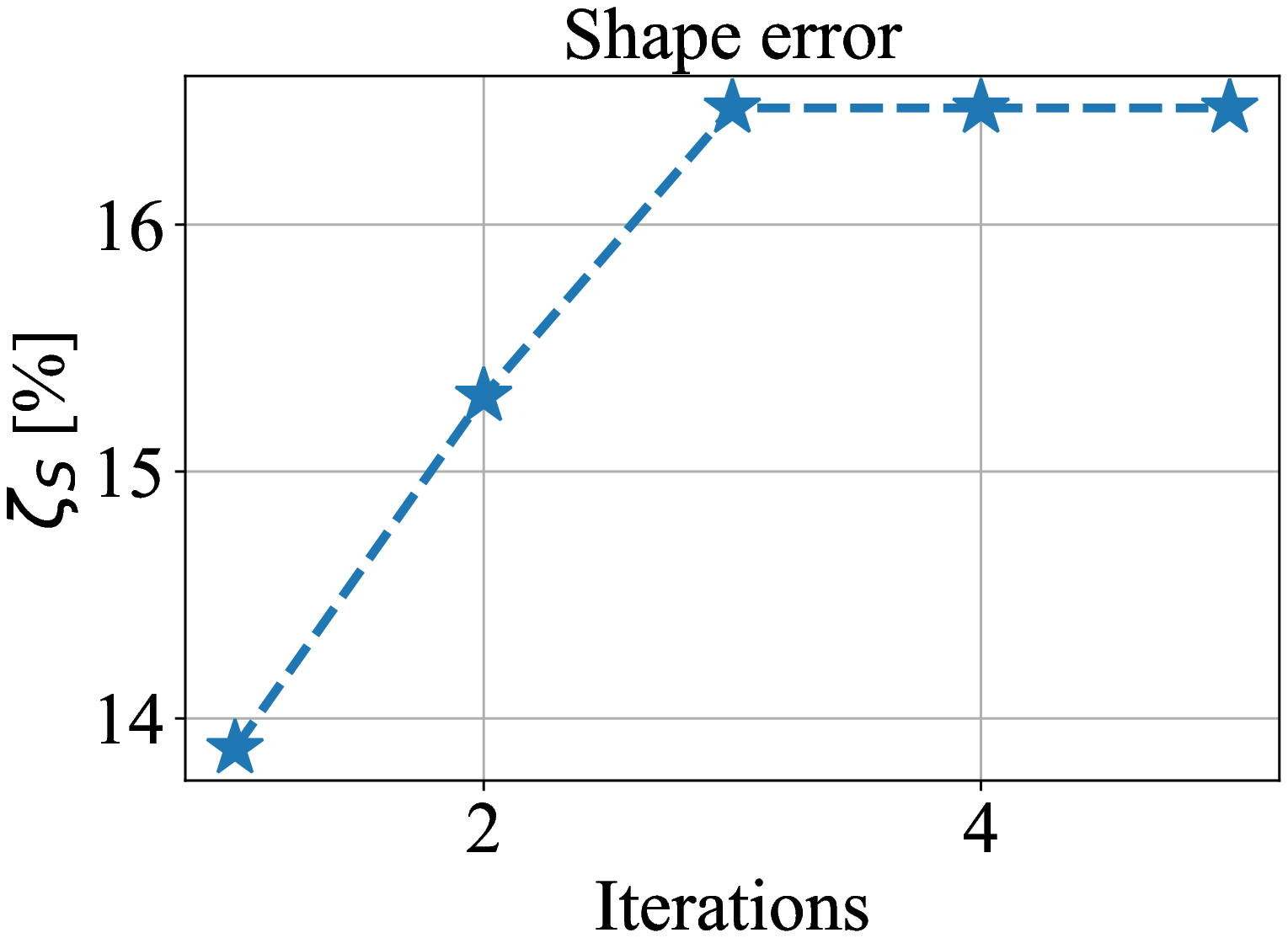}\label{fig:manual:results:zeta_p}}
		\caption{Results of manual definition usage example: (a) recovered image, (b) $\zeta_{RN}$, (c) $\zeta_{RPAD}$, (d) $\zeta_{\epsilon PAD}$, (e) $\zeta_{\epsilon OE}$, (f) $\zeta_{\epsilon BE}$, (g) $\zeta_{S}$, and (h) $\zeta_{P}$.}
		\label{fig:manual:results}
	\end{figure*}

	
	The recovered image and the graphics generated by the previous routine can be seen in Fig.\ref{fig:manual:results}. Note that the reconstructed image resembles that of Fig.\ref{fig:manual:scatterer}, both in shape and position and in contrast value. Convergence graphs show a regular decay of the indicators regarding the error both in contrast estimation and residuals of the equations. The execution time was: 18.6s\footnote{Computer configuration: macOS 10.15.7, 2.5 GHz Dual-Core Intel Core i5, 16GB 1600 MHz DDR3 RAM.}.

	Special attention needs to be given to the results of the two new indicators introduced in this work, i.e., $\zeta_S$ and $\zeta_P$. The graphs of these two indicators (Fig.\ref{fig:manual:results:zeta_s} and \ref{fig:manual:results:zeta_p}) show slight variations in the beginning (2 and 0.04\%, respectively) and constancy in the last three iterations (similarly to the other indicators). It is also noteworthy that $\zeta_S$ increased in the first iterations, suggesting that, by improving the contrast estimation, the method loses some of the scatterer contours. Therefore, these two indicators suggest a difficulty of the method in promoting large variations in the shape and position of the scatterers. In other words, the results of these indicators suggest that the initial solution is already of good quality and that the method's iterations have little effectiveness in improving the image in terms of shape and position. Furthermore, the final values of these indicators are information that suggests the application of the method in situations where errors in shape and position of less than 20 and 1\%, respectively, are required. Therefore, the two new metrics presented in this work are relevant because they allow: (i) to indicate the method's ability to promote improvements in the shape and position of the scatterers throughout the iterative process; and (ii) to quantify the error, thus producing information that can serve as a reference and support the application of methods in specific situations. Thus, the introduction of these new indicators contributes to the literature as a simpler and more general way to quantify the error in these two important aspects of reconstruction, i.e., shape and position.
	
	\subsection{Case Study}\label{sub:casestudy}
	
	In this example, a case study will be defined and run. Then, the results will be analyzed. For this example, the following settings are considered: (i) $\lambda_b = $ 0.5 (m); (ii) $N_S =$ 10, $N_M =$ 9; (iii) $L_x = L_y =$ 0.8 (m) (1.6 $\lambda_b$); (iii) $R_O = $ 1 (m) (2$\lambda_b$); (iv) $\epsilon_{rb} = 4$, $\sigma_b=0$ (S/m); and (v) $E_0 = $ 1 (V/m). These settings are similar to \cite{salucci2017multifrequency}, however, incident plane waves and only perfect dielectric materials assumptions are considered.
	
\begin{figure}
		\centering
		\includegraphics[width=3.5in]{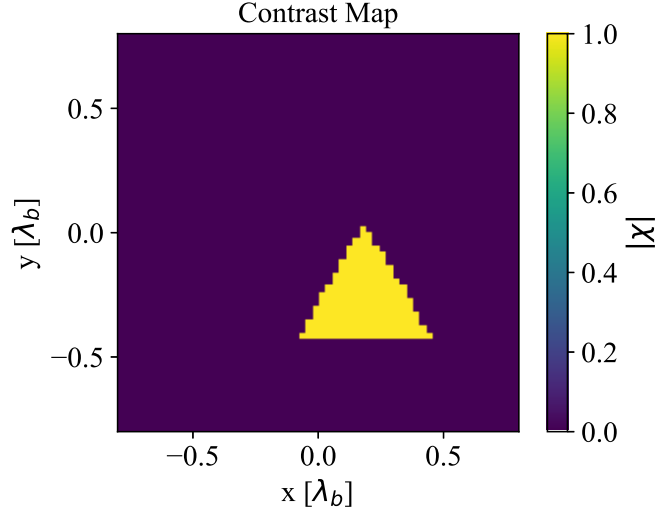}
		\caption{Image of the scatterer used in the case study example.}
		\label{fig:casestudy:scatterer}
	\end{figure}

	
	For this test, a single scatterer, whose geometry is an equilateral triangle, is addressed (Fig.\ref{fig:casestudy:scatterer}). Its length has been set to fit a circle with a radius of 0.16 (m) (0.32$\lambda_b$) and its contrast has been set to 1. To synthesize the scattered field data, the following definitions have been considered: (i) the resolution of the ground-truth image has been set to 60$\times$60 pixels (ii) the same MoM-CG-FFT version of the previous example is used, and (iii) noise of 1 (\%/sample) was added.
	
	In this study, only EAs are addressed. The settings for the three algorithms are listed below:
	
	\begin{enumerate}
		\item DE/rand/2/bin: $F=$ 0.5 (Mutation Factor), $CR=$ 0.5 (Crossover Rate), Binary Tournament Selection \cite{rocca2011differential};
		\item PSO: $c_1=c_2=$ 2.0 (acceleration coefficients), $w=$ 0.4 (inertia weight) \cite{salucci2017multifrequency}.
		\item GA: Simulated Binary Crossover ($\eta =$ 5, crossover probability $=$ 100\%), Polynomial Mutation ($\eta = $ 5, mutation probability $=$ 100\%), Binary Tournament Selection \cite{deb2001self}.
	\end{enumerate}
	
	The following settings will be considered for all algorithms: (i) Population size: 250; (ii) Representation: \textit{DiscretizationElementBased}; (iii) Objective-function: \textit{WeightedSum}; (iv) Initialization: \textit{BornApproximation}; (v) Number of executions per algorithm: 30; (vi) Range for contrast variable values: [0, 1]; (vii) Range for real and imaginary values of total field variables: [-5, 5]; (viii) Discretization: 7$\times$7 elements (539 variables); (ix) stop criterion: 5,000 iterations. The implementation of the methods with these settings can be seen in the following code:
	
	{\footnotesize 
\begin{lstlisting}[language=Python]
import stochastic as stc
import evolutionary as evo
import stopcriteria as stp
import result as rst
import richmond as ric
import mom_cg_fft as mom

from evoalglib import initialization as ini
from evoalglib import objectivefunction as obj
from evoalglib import representation as rpt
from evoalglib import boundary as bc
from evoalglib import selection as slc
from evoalglib import crossover as cross
from evoalglib import mutation as mut
from evoalglib import de
from evoalglib import pso
from evoalglib import ga

population_size = 250
variables_per_dimension = 7
contrast_max = 1.
total_field_max = 5.

initialization = ini.BornApproximation()
objfun = obj.WeightedSum()
representation = rpt.DiscretizationElementBased(
    ric.Richmond(
        configuration,
        variables_per_dimension,
        state=True
    ),
    contrast_max,
    total_field_max
)
stopcriterium = stp.StopCriteria(
    max_iterations=1000
)
outputmode = stc.OutputMode(
    stc.EACH_EXECUTION
)

DE = evo.EvolutionaryAlgorithm(
    population_size,
    initialization,
    objfun,
    representation,
    de.DifferentialEvolution(
        bc.Reflection(),
        slc.BinaryTournament(
            elitism=True,
            pair_selection='random'
        ),
        de.RAND,
        .5, # mutation factor
        cross.Binomial(.5),
        pcross=1,
        index_selection='random'),
    stopcriterium,
    outputmode,
    alias='de',
    parallelization=True,
    forward_solver=mom.MoM_CG_FFT()
)

PSO = evo.EvolutionaryAlgorithm(
    population_size,
    initialization,
    objfun,
    representation,
    pso.ParticleSwarmOptimization(
        bc.Reflection(),
        acceleration=2.,
        inertia=.4
    ),
    stopcriterium,
    outputmode,
    alias='pso',
    parallelization=True,
    forward_solver=mom.MoM_CG_FFT()
)

GA = evo.EvolutionaryAlgorithm(
    population_size,
    initialization,
    objfun,
    representation,
    ga.GeneticAlgorithm(
        bc.Reflection(),
        cross.SimulatedBinary(5.),
        1., # crossover probability
        mut.Polynomial(5),
        1., # mutation probability
        slc.BinaryTournament()
    ),
    stopcriterium,
    outputmode,
    alias='ga',
    parallelization=True,
    forward_solver=mom.MoM_CG_FFT()
)
\end{lstlisting}}
	
	The construction of the object of class \textit{CaseStudy} can be seen in the following code:
	
	{\footnotesize 
\begin{lstlisting}[language=Python]
mystudy = cst.CaseStudy(
    'mycasestudy',
    method=[DE, PSO, GA],
    discretization=discretization,
    test=inputdata,
    stochastic_runs=30,
    save_stochastic_runs=True
)
\end{lstlisting}}
	
\begin{figure*}
		\centering
		\subfloat[]{\includegraphics[width=.25\textwidth]{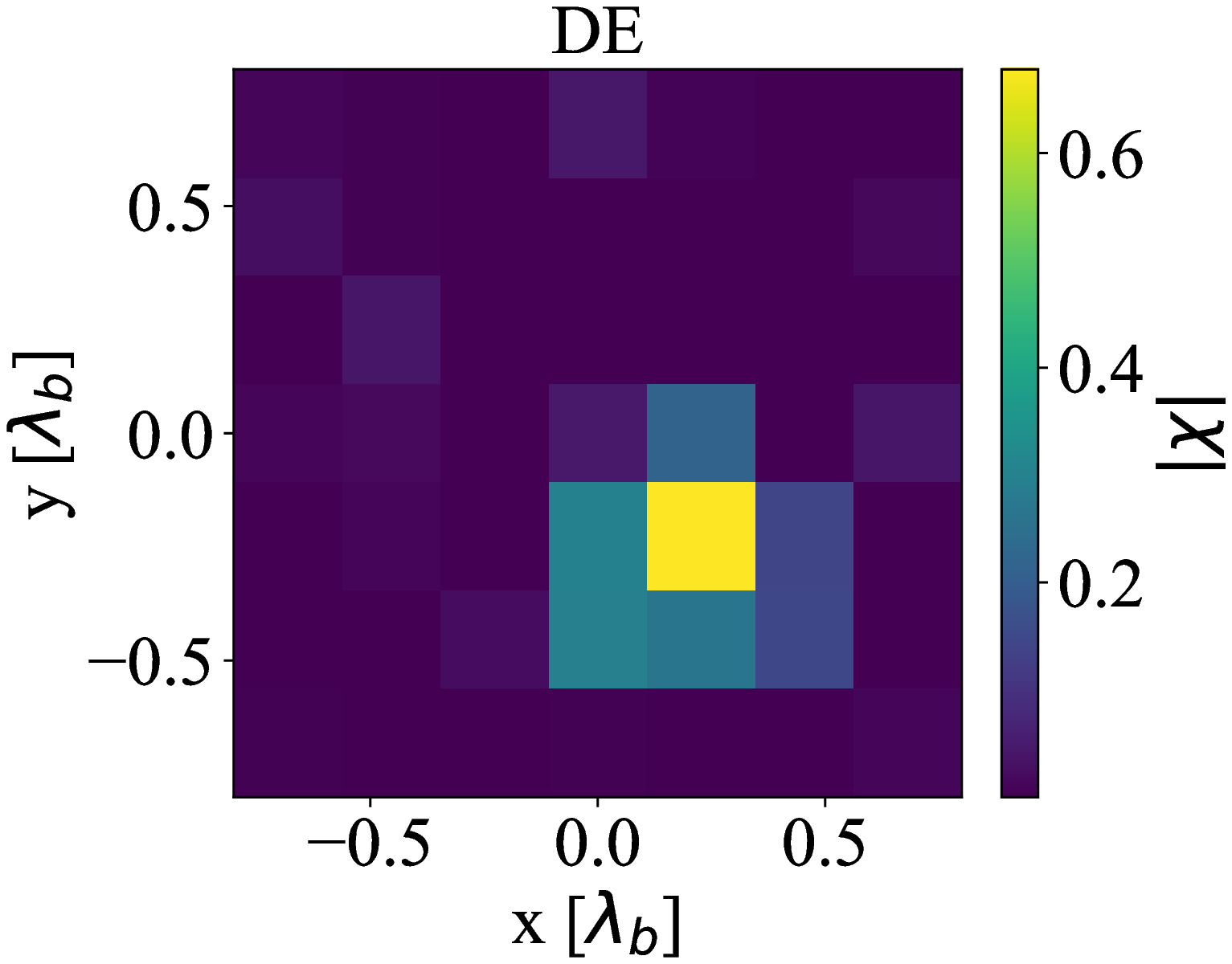}\label{fig:casestudy:results:rec:de}}
		\subfloat[]{\includegraphics[width=.25\textwidth]{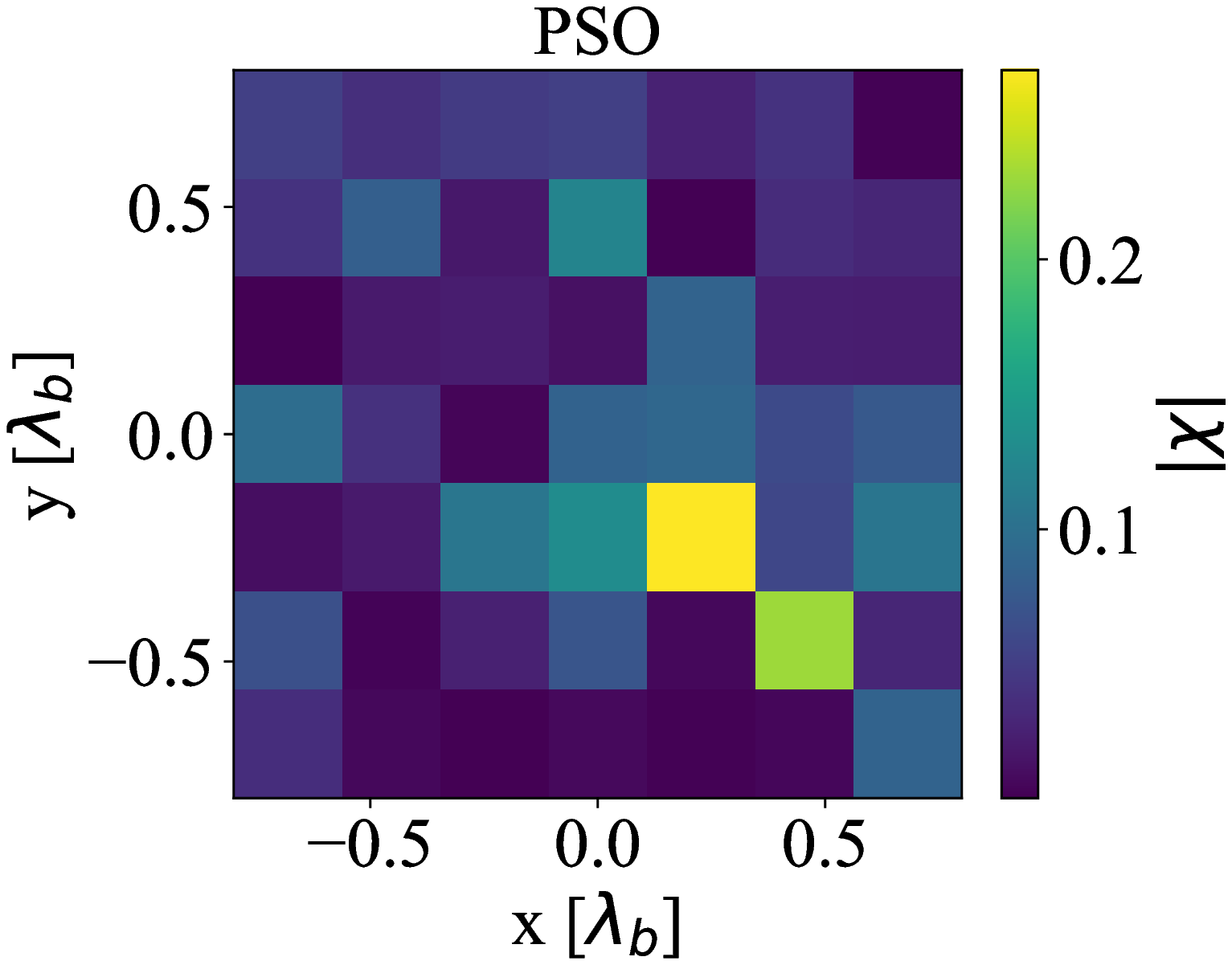}}
		\subfloat[]{\includegraphics[width=.25\textwidth]{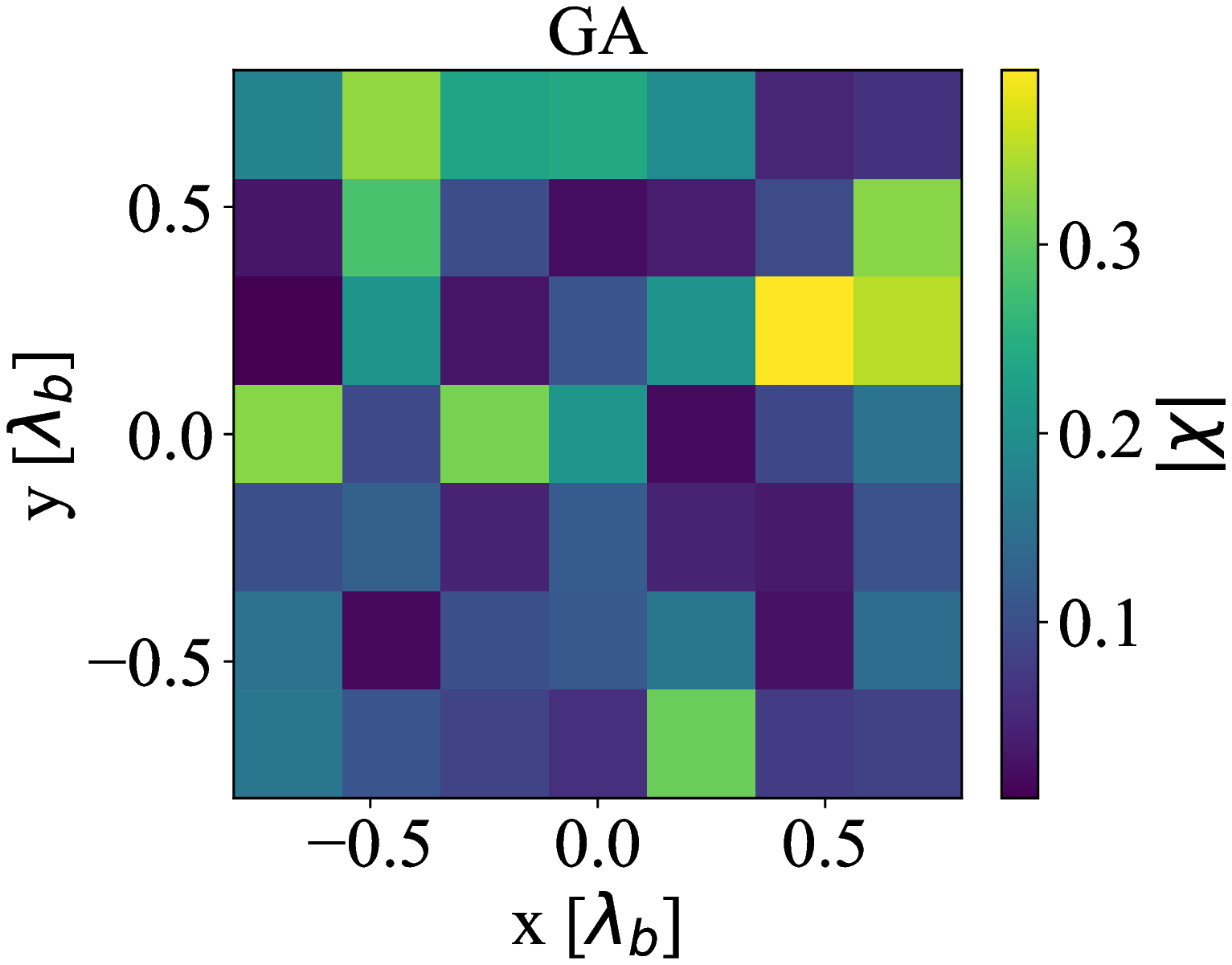}\label{fig:casestudy:results:rec:ga}}
		\subfloat[]{\includegraphics[width=.25\textwidth]{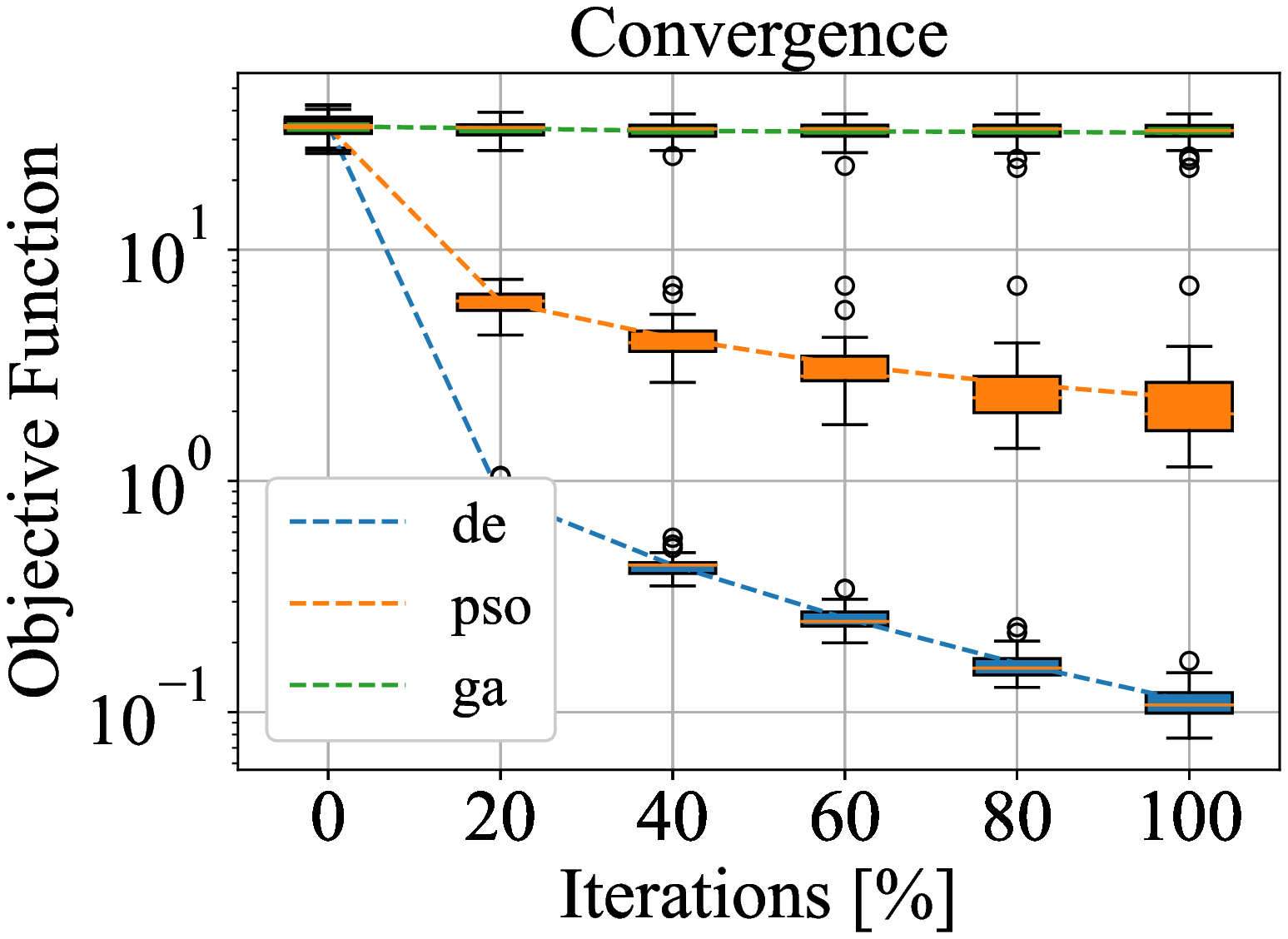}\label{fig:casestudy:results:conv}} \\
		\subfloat[]{\includegraphics[width=.25\textwidth]{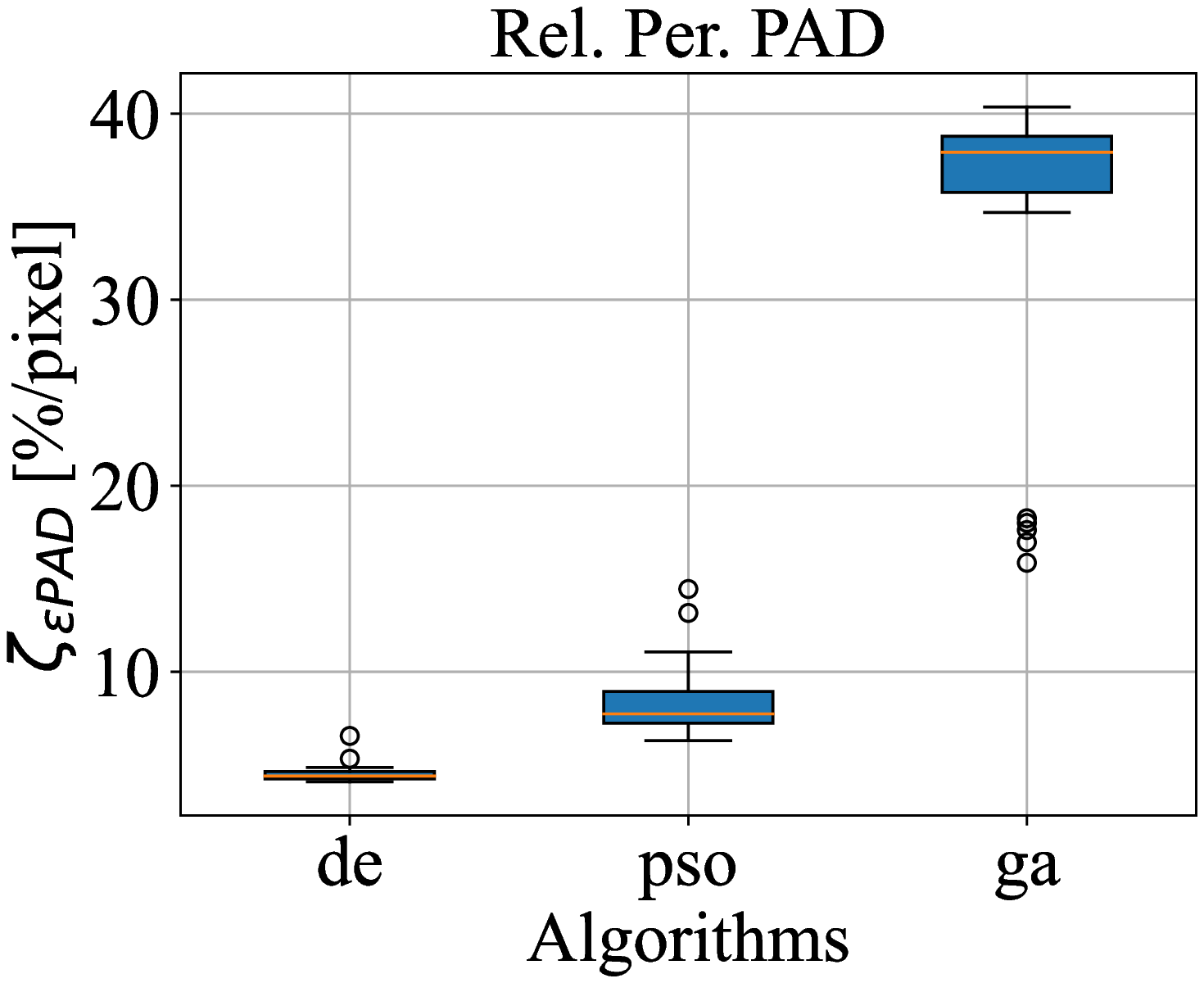}\label{fig:casestudy:results:boxplot}}
		\subfloat[]{\includegraphics[width=.25\textwidth]{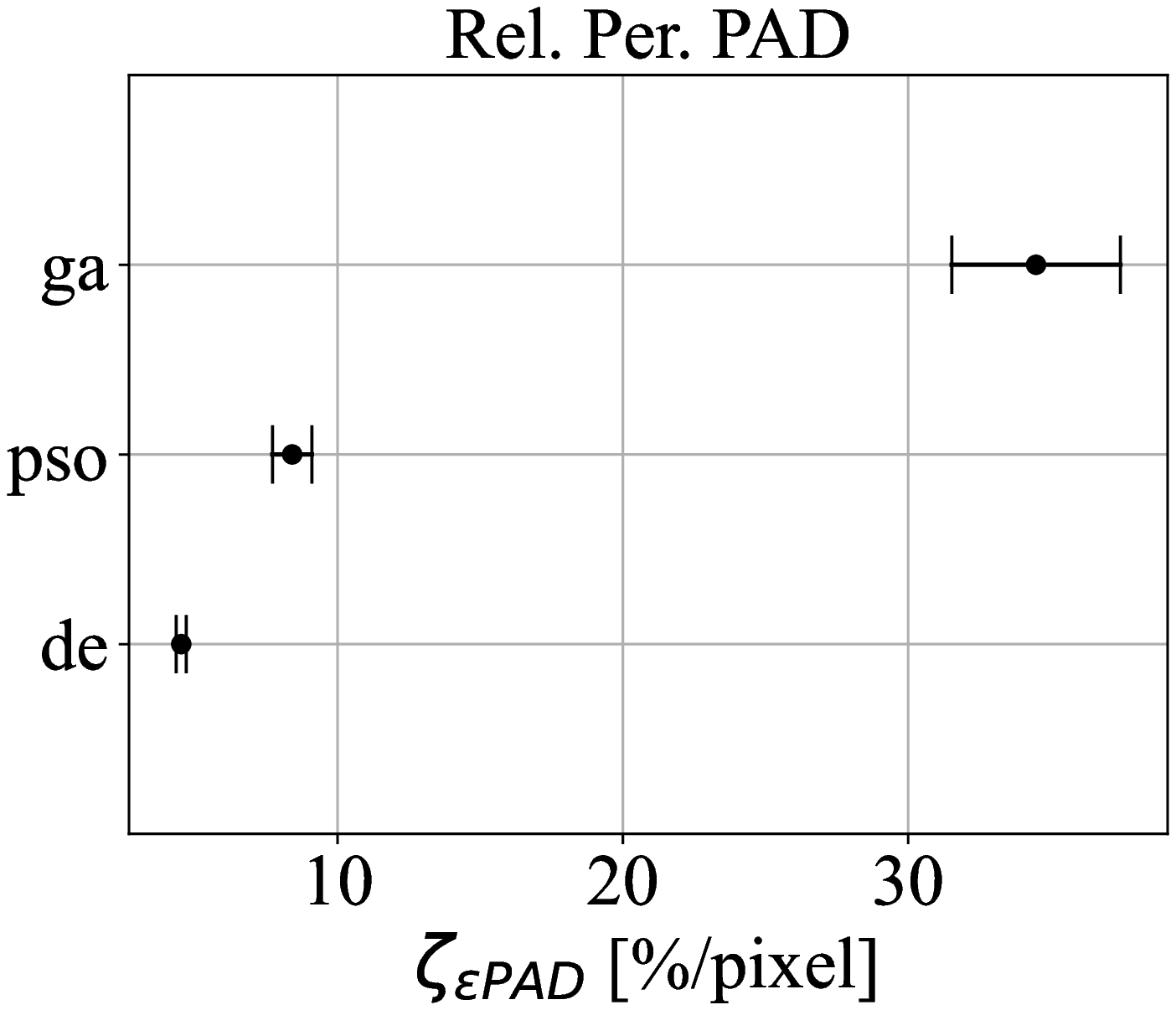}\label{fig:casestudy:results:confint}}
		\caption{Results obtained in the case study usage example: best image (according to $\zeta_{\epsilon PAD}$) recovered by (a) DE, (b) PSO, and (c) GA; (d) convergence of the $\zeta_{\epsilon PAD}$ indicador by the algorithms; (e) boxplot of final $\zeta_{\epsilon PAD}$ values obtained by the algorithms; (f) 95\% confidence interval for the average of the $\zeta_{\epsilon PAD}$ indicator for each algorithm.}
		\label{fig:casestudy:results}
	\end{figure*}

	Once the methods and input data are already entered, the object can perform the executions. Once the executions are carried out, the results can be analyzed through the routines implemented in the class. Fig.\ref{fig:casestudy:results} presents some of the possible graphics for this case study.
	
	The best images, according to the final value of the $\zeta_{\epsilon PAD}$ indicator, obtained by each algorithm (Fig. \ref{fig:casestudy:results:rec:de}-\ref{fig:casestudy:results:rec:ga}) suggest that DE had less difficulty locating and estimating the scatterer's contrast. While its best image shows a region whose contrast clearly stands out in relation to its surroundings, the same is not observed in the best image of the other algorithms.
	
	The objective-function convergence in DE runs is significantly different from the others (Fig. \ref{fig:casestudy:results:conv}). In addition to reaching a lower value faster, the variation in each iteration percentage is much lower. This observation suggests a uniformity in the DE search process among its executions, i.e., the algorithm tends to converge to the same region in the search space whenever it is executed.
	
	
	The final value of the $\zeta_{\epsilon PAD}$ indicator found by DE and PSO tends to be lower than GA (Fig.\ref{fig:casestudy:results:boxplot}). When the comparison routine between the three algorithms is applied, the result detects a difference in the performance of this indicator (p-value of Kruskal-Wallis H-Test: 7.198$\times$10$^{-18}$) and, in multiple comparisons, differences are detected in all pairs. Regarding the two pairs involving GA, the p-value of the Mann-Whitney U test was 3.2$\times$10$^{-11}$ in both cases. Regarding the pair DE-PSO, the p-value was  3.690$\times$10$^{-11}$. Although the normality assumption was not able to be assured for each mean, the 95\% confidence interval of the means plot (Fig.\ref{fig:casestudy:results:confint}) shows a slight difference in performance in this study between DE and PSO (at least 3 \%/pixel) where as, between DE and GA, the difference was more than 20 (\%/pixel). Therefore, the results indicate that the best performance in this case study, according to the $\zeta_{\epsilon PAD}$ indicator, was accomplished by DE (95\% confidence level).
	
	\subsection{Benchmark}\label{sub:benchmark}
	
	In this usage example, the \textit{Benchmark} class will be used to perform a comparative study between BIM, Born Approximation (BA), and DE. The same domain configuration is considered as in Subsection \ref{sub:casestudy}. Only one test set with the following specifications will be used: (i) contrast of each scatterer = 1; (ii) fixed radius per scatterer = 0.32$\lambda_b$; (iii) one scatterer per image; (iv) random polygons; (v) resolution of each image = 100$\times$100 pixels; (vi) 30 tests; (vii) 1 (\%/sample) noise; and (viii) the indicators $\zeta_{\epsilon PAD}$ and $\zeta_{\epsilon OE}$ will be addressed. The test suite implementation is seen below:
	
	{\footnotesize 
\begin{lstlisting}[language=Python]
import testset as tst
import experiment as exp
import result as rst
import mom_cg_fft as mom
mytestset = tst.TestSet(
    name='mytestset',
    configuration=configuration,
    contrast=1.,
    object_size=.32,
    resolution=(100, 100),
    density=None,
    map_pattern=exp.RANDOM_POLYGONS_PATTERN,
    sample_size=30,
    noise=1.,
    indicators=rst.REL_PERMITTIVITY_PAD_ERROR,
    contrast_mode=exp.FIXED_CONTRAST,
    object_size_mode=exp.FIXED_SIZE,
    density_mode=exp.SINGLE_OBJECT
)
mytestset.randomize_tests()
mytestset.generate_field_data(
    solver=mom.MoM_CG_FFT(
        tolerance=1e-3,
        maximum_iterations=5000
    )
)
\end{lstlisting}}
	
	The average DNL of the generated tests was approximately 0.1467. The same BIM and DE versions from the previous examples will be used again. BA will be equipped with the same regularizer as the BIM. Instead of 5,000 iterations, each DE execution will run now 10,000. With these settings, we can initialize and run the study:
	
	
	{\footnotesize 
\begin{lstlisting}[language=Python]
import benchmark as bmk
mybenchmark = bmk.Benchmark(
    'mybenchmark',
    method=[BA, BIM, DE],
    discretization=discretization,
    testset=mytestset
)
mybenchmark.run()
\end{lstlisting}}
	
\begin{figure*}
		\centering
		\subfloat[]{\includegraphics[width=.25\textwidth]{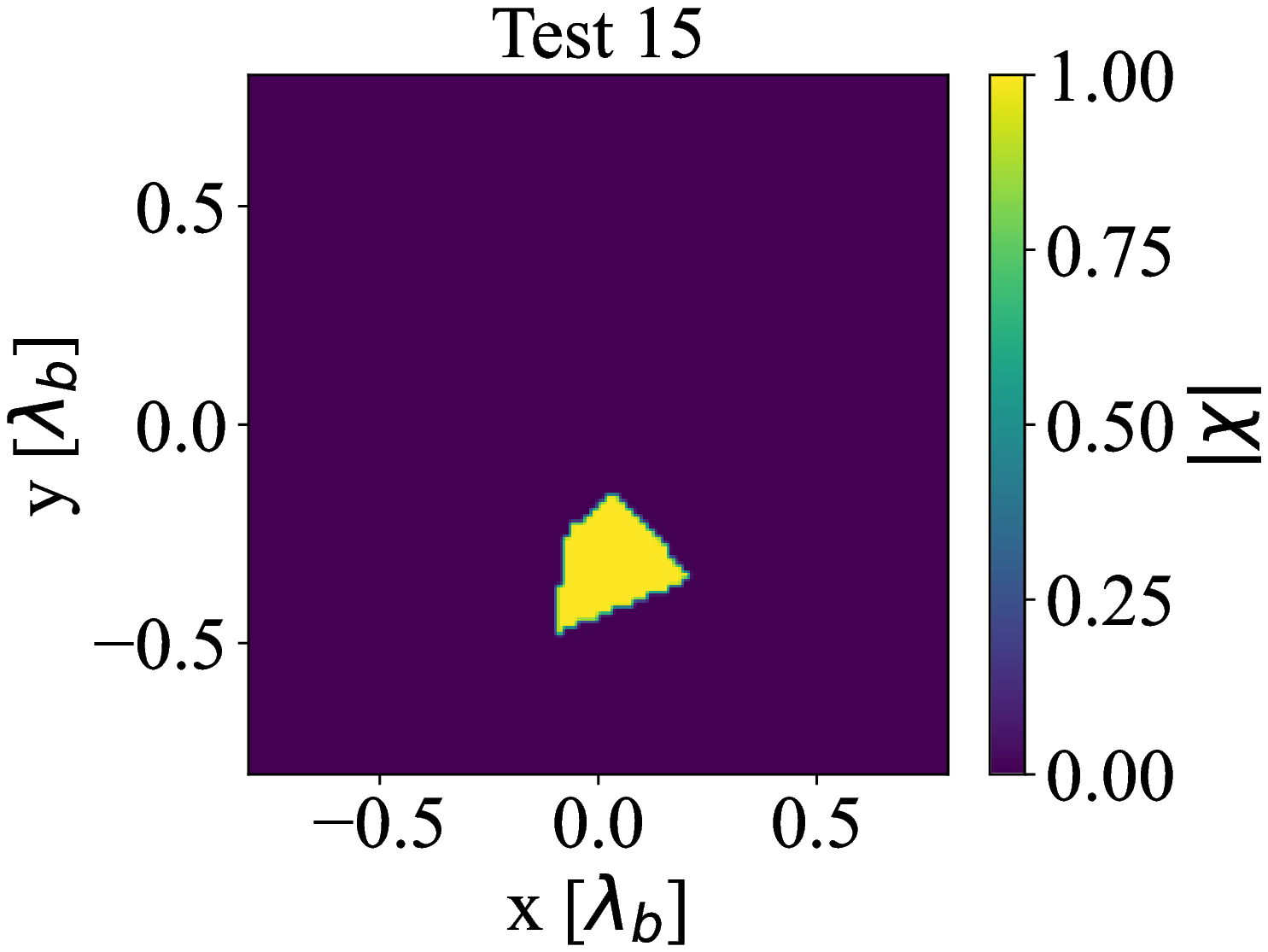}}
		\subfloat[]{\includegraphics[width=.25\textwidth]{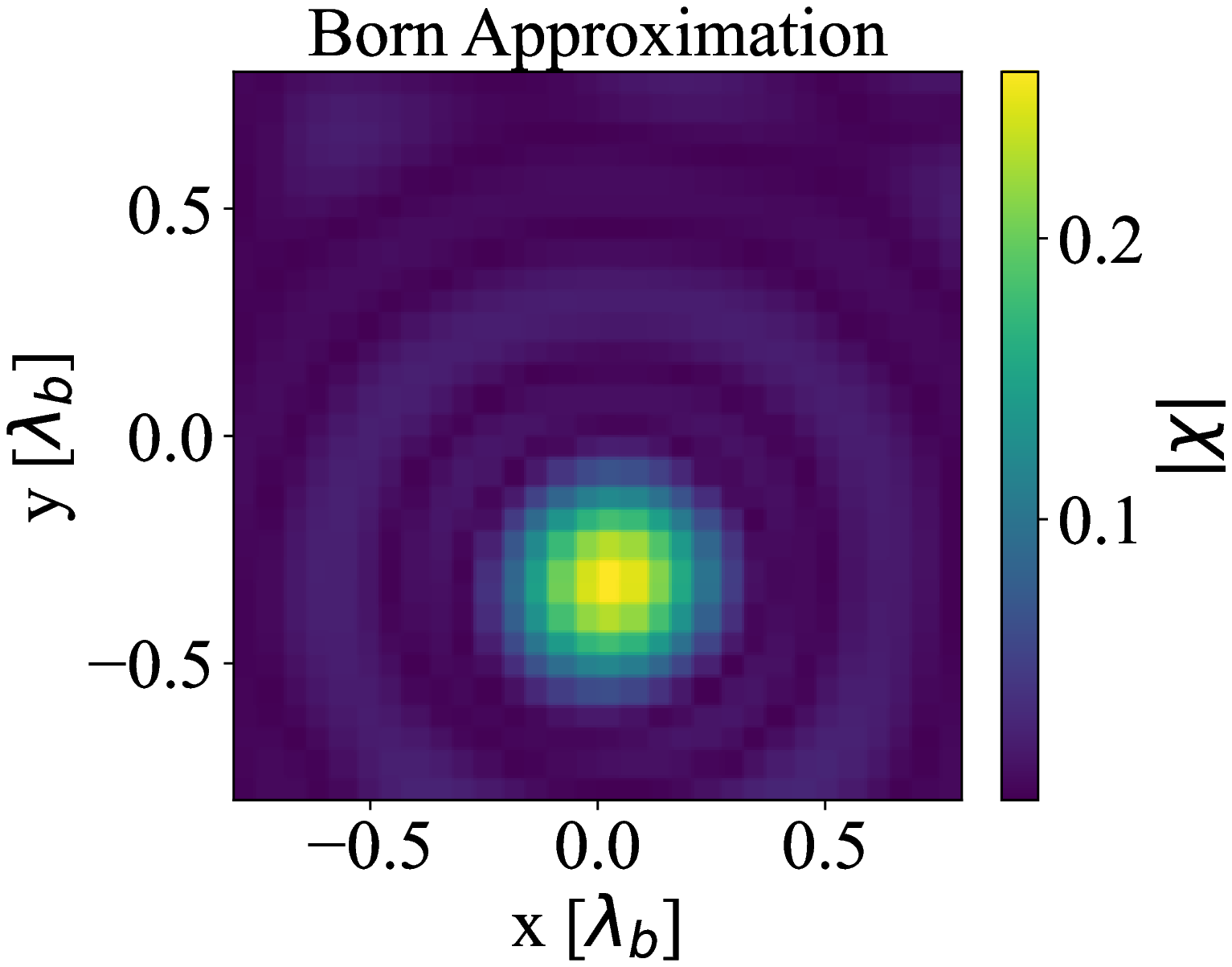}}
		\subfloat[]{\includegraphics[width=.25\textwidth]{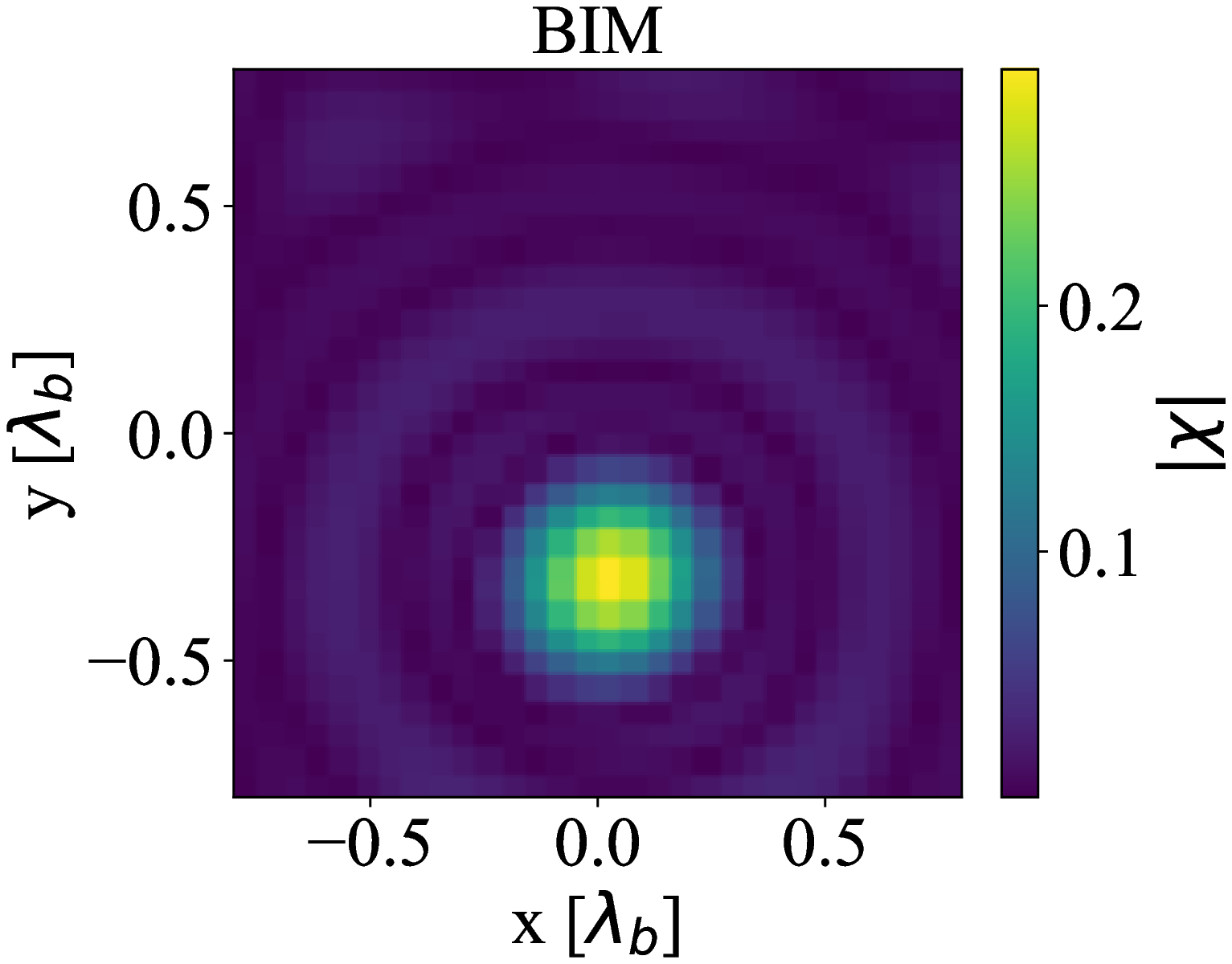}}
		\subfloat[]{\includegraphics[width=.25\textwidth]{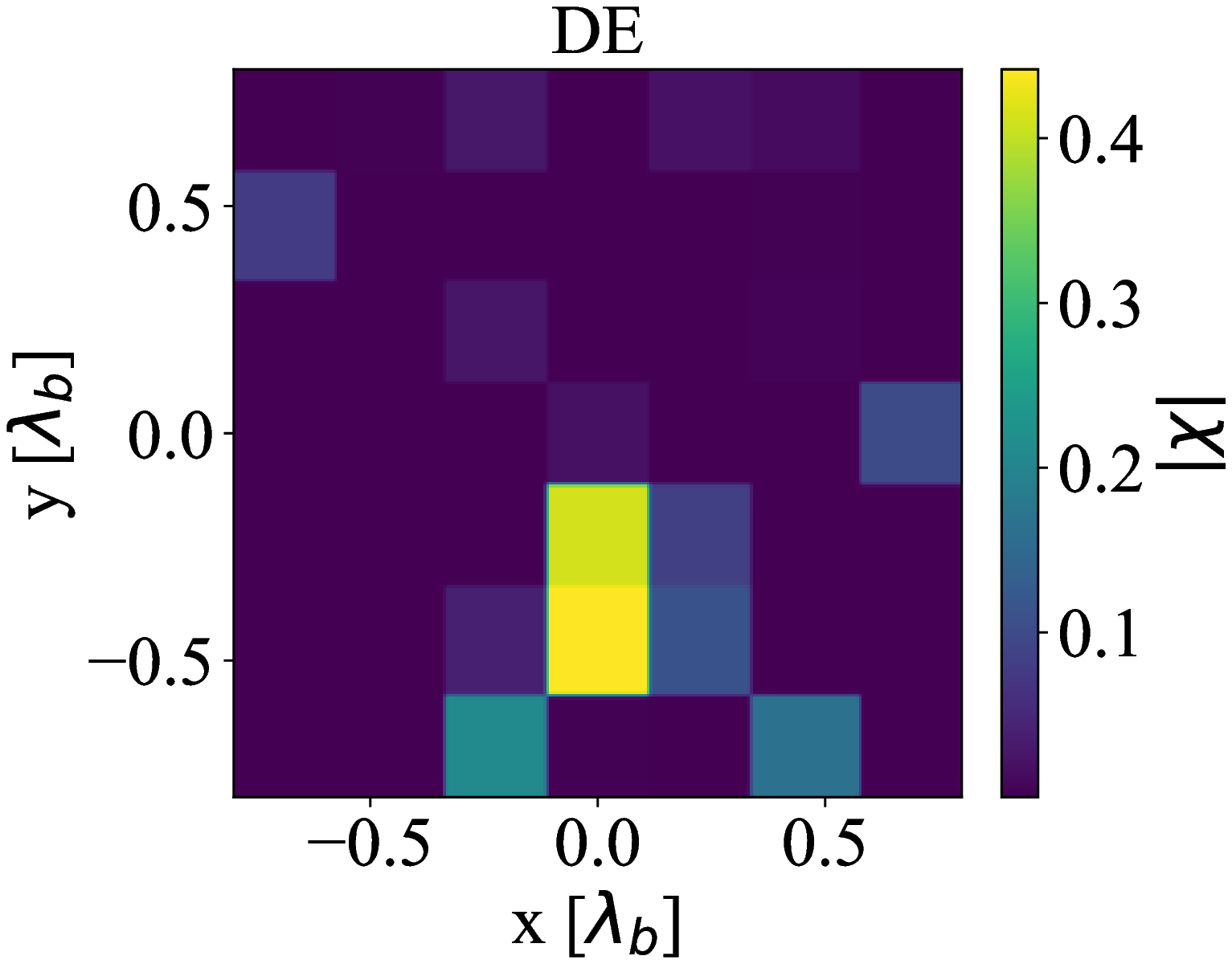}}
		\caption{Best images, according to $\zeta_{\epsilon PAD}$ indicator, recovered by each algorithm in the benchmark usage example. All algorithms achieved the lowest $\zeta_{\epsilon PAD}$ error in the test 16. The figure shows (a) the test 16 and (b) BA, (c) BIM, and (d) DE reconstructions.}
		\label{fig:benchmark:recons}
	\end{figure*}

	
	The best reconstructions of each algorithm, according to the lowest $\zeta_{\epsilon PAD}$ indicator achieved by each of them, are shown in Fig.\ref{fig:benchmark:recons}. Coincidentally, the three algorithms achieved the lowest $\zeta_{\epsilon PAD}$ error in the 16\textsuperscript{th} test. Therefore, the figure shows the test and the corresponding reconstructions done by each algorithm. The images from the deterministic methods are very similar, whereas DE made a better estimation of the contrast value.
	
	
\begin{figure*}
		\centering
		\subfloat[]{\includegraphics[width=.25\textwidth]{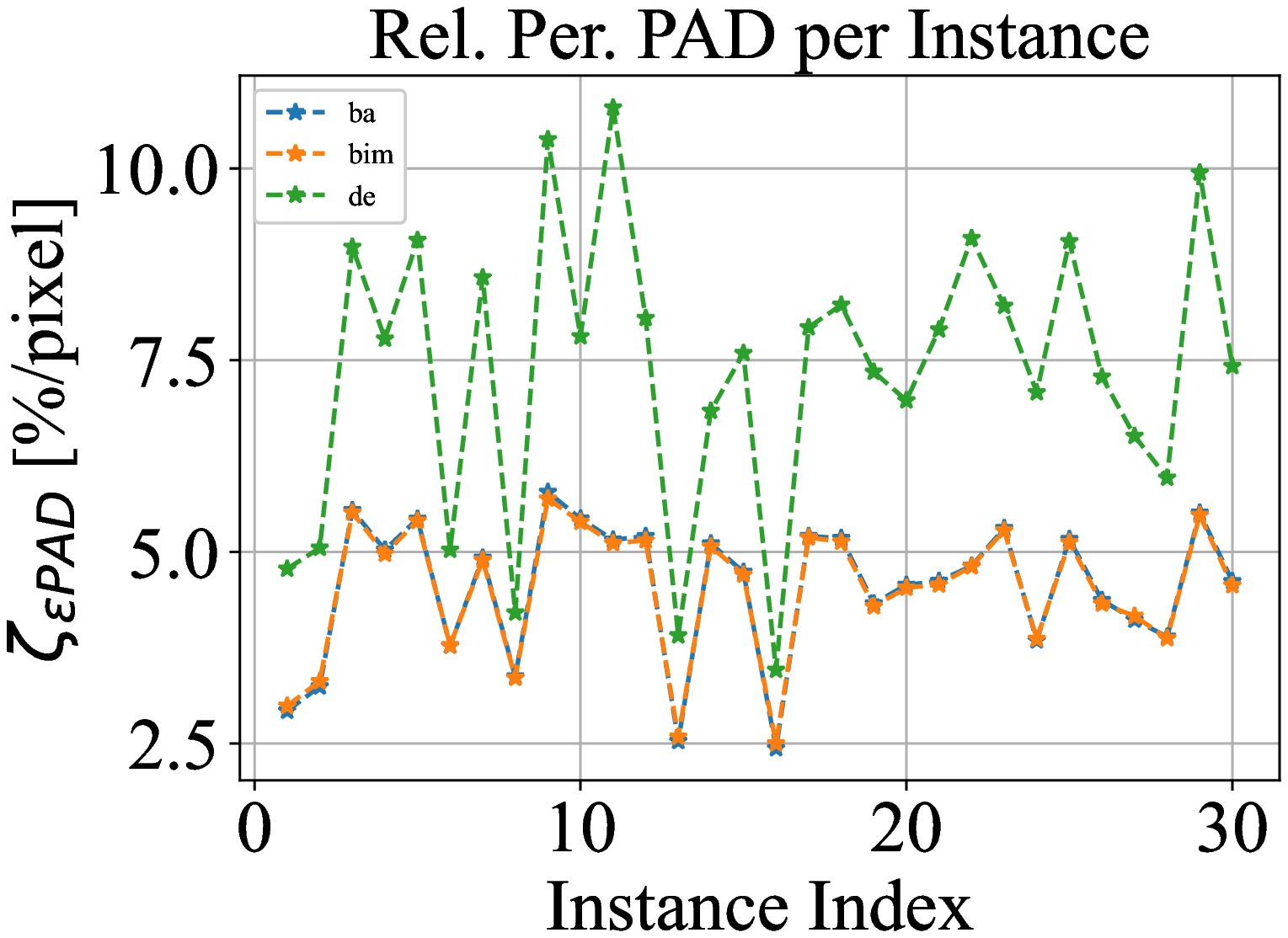}\label{fig:benchmark:others:plot}}
		\subfloat[]{\includegraphics[width=.25\textwidth]{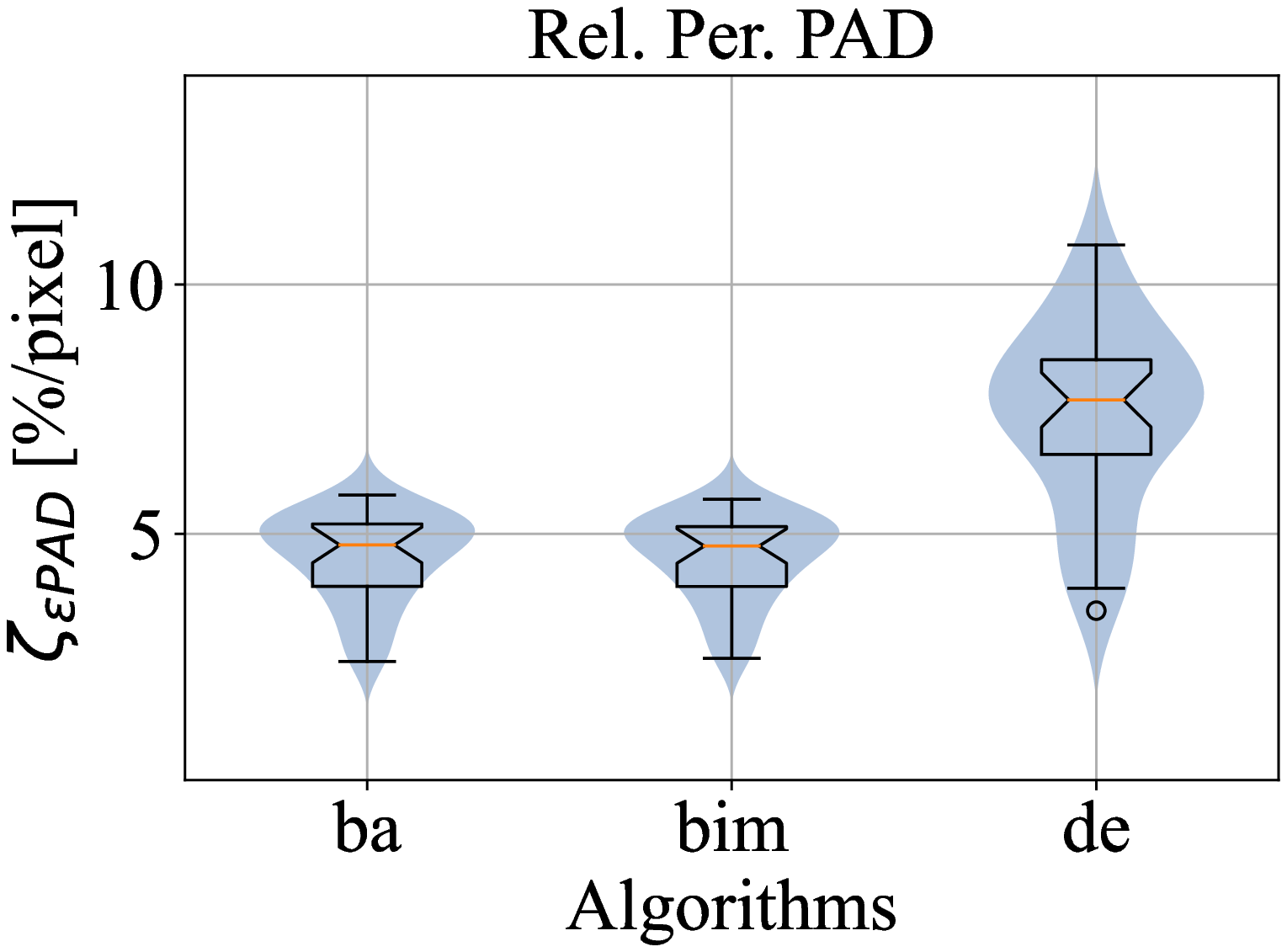}\label{fig:benchmark:others:violinplot}}
		\subfloat[]{\includegraphics[width=.25\textwidth]{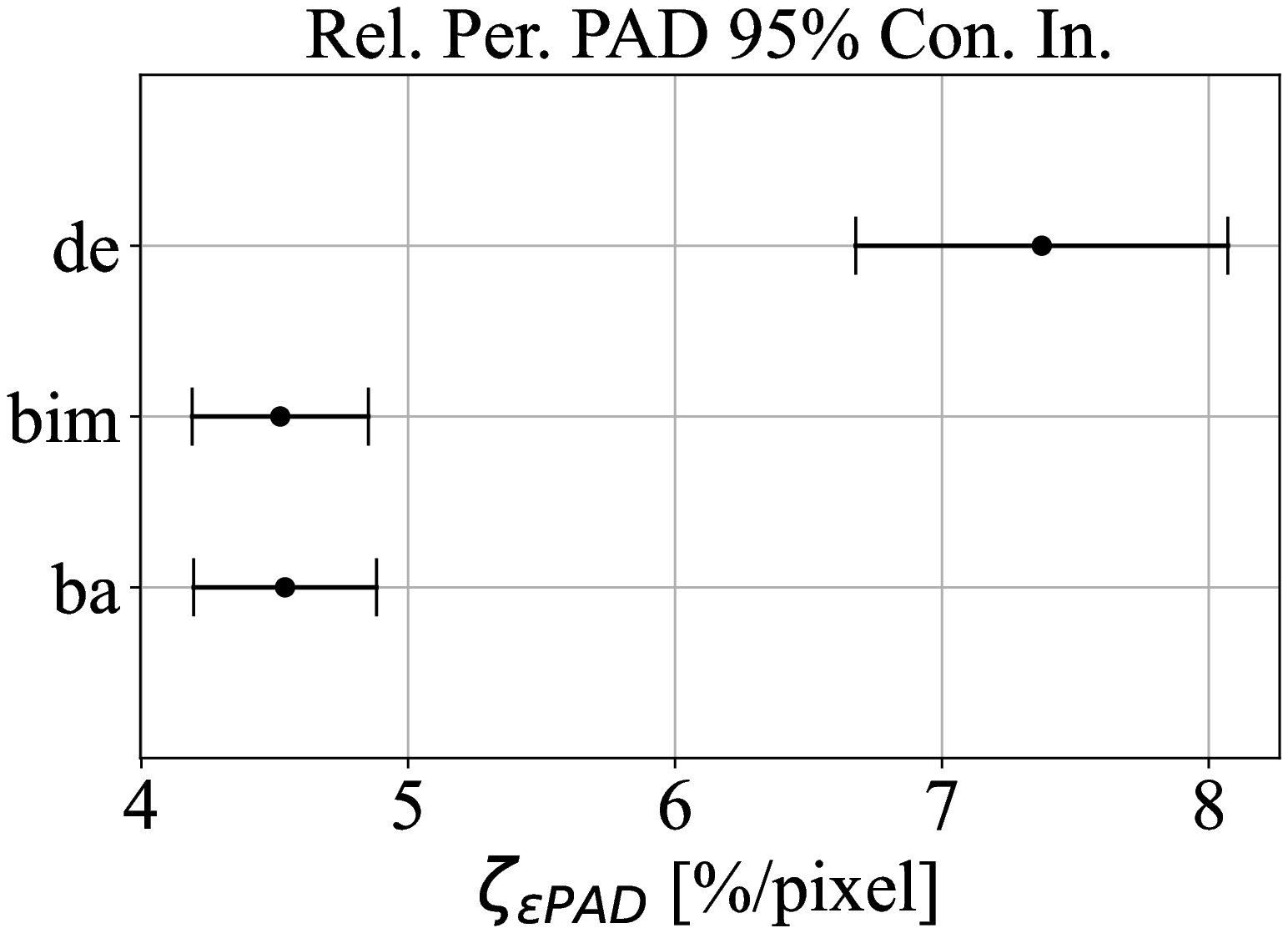}\label{fig:benchmark:others:confint}}
		\subfloat[]{\includegraphics[width=.25\textwidth]{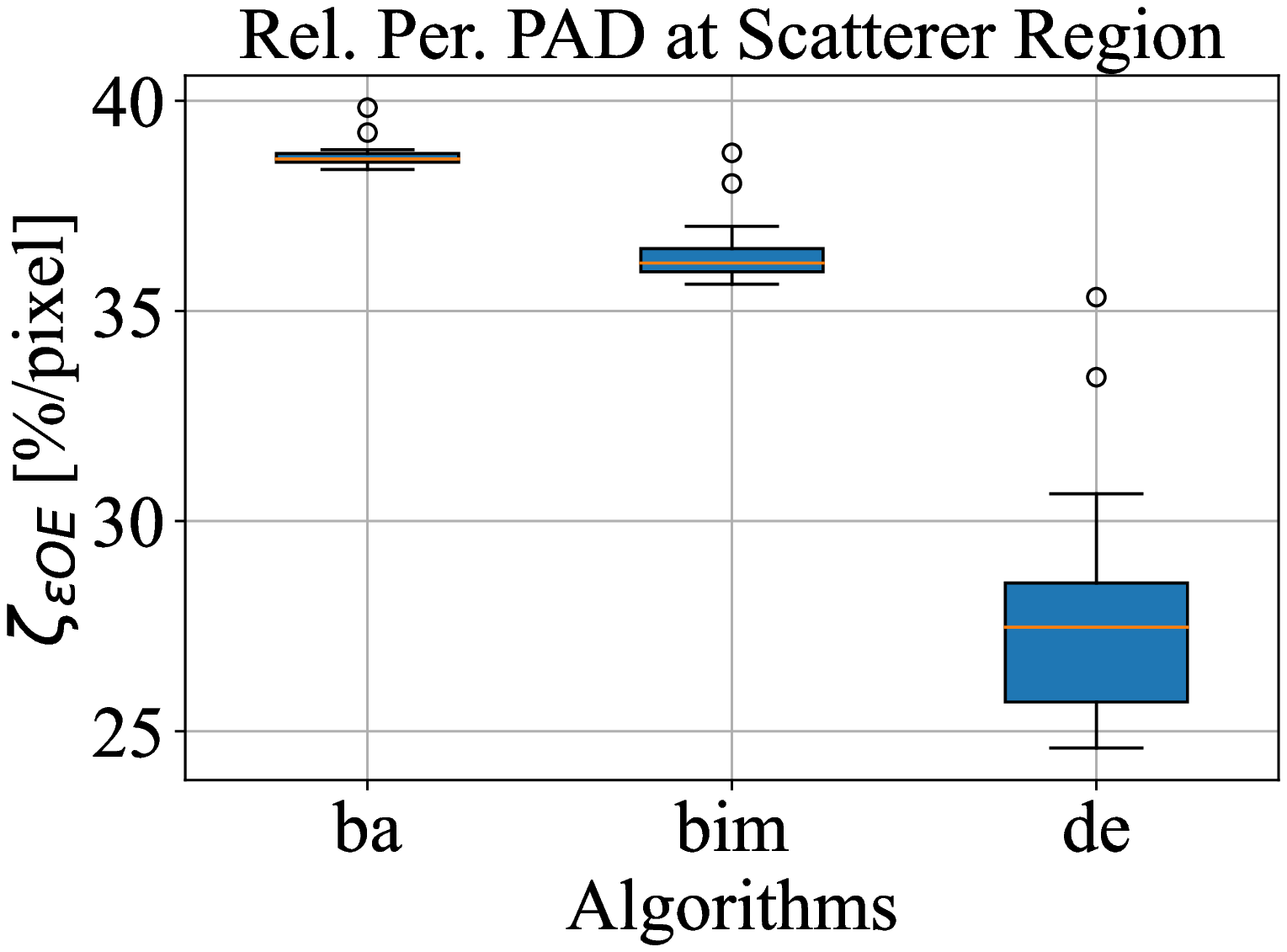}\label{fig:benchmark:others:boxplot}}
		\caption{Benchmark usage example results: (a) final $\zeta_{\epsilon PAD}$ value per instance;  (b) distribution of $\zeta_{\epsilon PAD}$ indicator; (d) 95\% confidence interval of $\zeta_{\epsilon PAD}$; (d) distribution of $\zeta_{\epsilon OE}$ indicator.}
		\label{fig:benchmark:others}
	\end{figure*}

	The results of the $\zeta_{\epsilon PAD}$ and $\zeta_{\epsilon OE}$ indicators are shown in Fig. \ref{fig:benchmark:others}. BIM obtained an $\zeta_{\epsilon PAD}$ value very close to the Born Approximation in each instance (Fig. \ref{fig:benchmark:others:plot}) and also when considering the distribution over the 30 tests (Fig.\ref{fig:benchmark:others:violinplot}). The performance of both algorithms seems to be different than DE and such difference is detected by Randomized Complete Block Design after the square-root data transformation (p-value: 5.686$\times10^{-32}$). Differences in performance are detected in multiple comparisons only when the following pairs are considered: BIM-DE (p-value: 4.702$\times10^{-17}$) and BA-DE (p-value: 2.679$\times10^{-17}$). The 95\% confidence interval plot shows a difference of, at least, 1.5 (\%/pixel) between the deterministic methods and the stochastic one. Although the performance of DE in the same indicator seems to be worse, the estimation of the contrast value of the scatterers seems to be better than the other algorithms (Fig\ref{fig:benchmark:others:boxplot}). Such an indication is confirmed when the performance of DE in the $\zeta_{\epsilon OE}$ indicator is compared against BA and BIM (p-value of Wilcoxon's test : 1.734$\times10^{-6}$, both cases).
	
	Summing up, the results allow the following inferences (95\% confidence level) about the performance of these three algorithms for the considered problem configuration: (i) there is no significant difference between BIM and BA when considering the average error on permittivity estimation in the whole image; (ii) although DE has a large average error when considering the whole image, it approximates better the relative permittivity of the scatterer.
	
	\section{Conclusion}\label{sec:conclusion}
	
	In this article, the \textbf{eispy2d} library is presented. It implements a set of tools to represent two-dimensional EISP instances, develop algorithms, test them and analyze their results. The library organizes problem entities into classes, following the Object-Oriented Programming paradigm. These classes attempt to embody the diversity of techniques and concepts in the literature in a broad fashion.
	
	The \textbf{eispy2d} library allows users to set up tests for the algorithms in addition to systematizing studies. From the specification of the problem domains and the nature of the incident field, the user can manually build a problem with the support of routines that insert scatterers in the image and forward solvers to synthesize the input data. Algorithms can be run manually or within a specific class for case studies, which has adequate tools for analyzing the results. On the other hand, if the objective is to measure the average performance and perform comparisons among different versions of the same algorithm or different methods, the library also provides a specific class for benchmark studies. This class has tools for (i) the generation of random problems organized in test sets following user specifications regarding the characteristics of the problem; (ii) measuring the performance of algorithms based on a set of quality indicators; (iii) the comparison of the average performance of the indicators among the algorithms with the support of adequate statistical tools.
	
	The purpose of library implementation is to provide a common platform for testing new ideas and save the researcher from having to implement the entire problem structure when interested in investigating a new technique. This work also proposes two new indicators for the literature, specific to measure the quality of location and reconstruction of scatterers' shapes. Furthermore, as far as the authors are aware, it is the first time in the literature of techniques for EISPs that an approach for estimating and comparing average performance supported by statistical tools has been developed. Therefore, this work contributes to the state-of-the-art literature of methods for EISPs by improving how these algorithms are studied.
	
	In the future, new tools for \textbf{eispy2d} are planned. In addition to implementing new forms of discretization (such as the use of triangular functions within the scope of the Collocation Method), other methods will be implemented (such as DBIM and CSI). We also intend to implement new structures to support different integral formulations, virtual experiments and a random test generation approach based on specifying a level of DNL. In addition, better techniques for defining the numbers of samples in test sets and repetitions of stochastic algorithms are also intended \cite{campelo2020sample}. Finally, we want to implement factor analysis as a new form of experimentation whose objective is to investigate the correlation and impact of different specifications factors (e.g., parameters of algorithms or scatterers) in the performance of the algorithms.

	\bibliographystyle{IEEEtran}
	\bibliography{IEEEabrv,refs.bib}

\end{document}